\documentclass[aps,prb,reprint,superscriptaddress,floatfix]{revtex4-1}

\usepackage{graphicx}
\usepackage{hyperref}
\usepackage{textcomp}
\usepackage{makecell}
\usepackage[english]{babel}
\usepackage{blindtext}
\usepackage{color}
\usepackage{soul}
\usepackage{placeins}
\usepackage{siunitx}
\usepackage{changes}

\definechangesauthor[name={Per cusse}, color=orange]{per}

\begin{document}

\title{Further investigations of the deep double donor magnesium in silicon}
\author{R. J. S. Abraham}
\author{A. DeAbreu}
\author{K. J. Morse}
\affiliation{Department of Physics, Simon Fraser University, Burnaby, British Columbia, Canada V5A 1S6}
\author{V. B. Shuman}
\affiliation{Ioffe Institute, Russian Academy of Sciences, 194021 St. Petersburg, Russia}
\author{L. МM. Portsel}
\affiliation{Ioffe Institute, Russian Academy of Sciences, 194021 St. Petersburg, Russia}
\author{АA. N. Lodygin}
\affiliation{Ioffe Institute, Russian Academy of Sciences, 194021 St. Petersburg, Russia}
\author{Yu. A. Astrov}
\affiliation{Ioffe Institute, Russian Academy of Sciences, 194021 St. Petersburg, Russia}
\author{N. V. Abrosimov}
\affiliation{Leibniz Institute for Crystal Growth, 12489 Berlin, Germany}
\author{S. G. Pavlov} 
\affiliation{Institute of Optical Sensor Systems, German Aerospace Center (DLR), 12489 Berlin, Germany}
\author{H.-W. H$\ddot{\text{u}}$bers}
\affiliation{Institute of Optical Sensor Systems, German Aerospace Center (DLR), 12489 Berlin, Germany}
\affiliation{Humboldt Universit$\ddot{\text{a}}$t zu Berlin, Department of Physics, 12489 Berlin, Germany}
\author{S. Simmons}
\affiliation{Department of Physics, Simon Fraser University, Burnaby, British Columbia, Canada V5A 1S6}
\author{M. L. W. Thewalt}
\email[Corresponding author: ]{thewalt@sfu.ca}
\affiliation{Department of Physics, Simon Fraser University, Burnaby, British Columbia, Canada V5A 1S6}

\date{\today}

\begin{abstract}
The deep double donor levels of substitutional chalcogen impurities in silicon have unique optical
properties which may enable a spin/photonic quantum technology. The interstitial magnesium impurity (Mg$_i$) in silicon is also a deep double donor but has not yet been studied in the same detail as have the chalcogens. In this study we look at the neutral and singly ionized Mg$_i$ absorption spectra in natural silicon and isotopically enriched 28-silicon in more detail. The 1s(A$_1$) to 1s(T$_2$) transitions, which are very strong for the chalcogens and are central to the proposed spin/photonic quantum technology, could not be detected. We observe the presence of another double donor (Mg$_{i*}$) that may result from Mg$_i$ in a reduced symmetry configuration, most likely due to complexing with another impurity. The neutral species of Mg$_{i*}$ reveal unusual low lying ground state levels detected through temperature dependence studies. We also observe a shallow donor which we identify as a magnesium-boron pair.
\end{abstract}

\maketitle
\FloatBarrier
\section{Introduction}
A recent proposal suggested the use of chalcogen deep double donors, such as selenium (Se) in silicon (Si), as the basis for a scalable qubit-photonic cavity technology \cite{Morse2017}. Like the chalcogens, the Group-II impurity magnesium is a helium-like deep double donor in silicon, though it is interstitial (Mg$_i$) rather than substitutional \cite{Franks1967, Ho1972}. Within Group-II this is unusual, with other impurities of that group, such as zinc and beryllium, forming substitutional double acceptors in silicon \cite{Carlson1957, Robertson1968, Crouch1972}, although there is evidence that some magnesium may indeed become substitutional (Mg$_s$) and act as a double acceptor \cite{Baber1988}. Many absorption features of interstitial magnesium's neutral (Mg$_i^{0}$) and ionized (Mg$_i^{+}$) species have been uncovered in past investigations \cite{Franks1967,Ho1972,Ho1993,Ho1998,Ho2003,2Ho2003,Ho2006}, although transitions from the 1s(A$_1$) ground state to s-like excited states, which are quite strong for the chalcogens \cite{Grimmeiss1981, Morse2017} and central to their potential use as photonically accessible spin qubits, have never been observed for Mg$_i$. \par

Early optical and electrical studies of Mg-diffused Si established Mg$_i$ as a double donor impurity \cite{Franks1967}.  The detailed absorption and piezoabsorption studies of Ho and Ramdas \cite{Ho1972} established the tetrahedral interstitial site for both Mg$_i^{0}$ and Mg$_i^{+}$, with ionization energies of 107.50 and 256.47 meV, respectively. These authors were also the first to comment on the absence of any visible 1s(A$_1$) to 1s(T$_2$) absorption for either Mg$_i^0$ or Mg$_i^+$, and to identify a doublet splitting of the 2p$_{\pm}$ transition for Mg$_i^{+}$ as resulting from a central-cell effect (for T$_d$ symmetry 2p$_{\pm}$ has the representation 2T$_1$+2T$_2$, and transitions from 1s(A$_1$) to T$_2$ states are dipole-allowed) \cite{Ho1972}. In an electrical and optical study of Mg-diffused Si, Lin \cite{Lin1982} detected a number of donor levels besides those of Mg$_i^0$ and Mg$_i^+$, with ionization energies of 40, 55, 80, and 93 meV. They also made the prescient suggestions that some of these unidentified species could result from complexes of Mg with other  rapidly diffusing impurities, such as transition metals, introduced during the Mg diffusion process, or from Mg in an interstitial site other than T$_d$. An earlier EPR study of Mg-diffused silicon by Baxter and Ascarelli had proposed that Mg might be photoconverted from the normal interstitial  T$_d$  site to a different interstitial site such as C$_{3v}$ \cite{Baxter1973}. A study using photothermal ionization spectroscopy found the same unidentified donor species with ionization energies of 55 and 93 meV, in addition to Mg$_i^0$ and Mg$_i^+$ \cite{Kleverman1986}. \par

Later studies undertaken by the Ho group \cite{Ho1993,Ho1998,Ho2003,2Ho2003,Ho2006} have included observations of a wider range of excited states for Mg$_i^0$ and Mg$_i^+$, resulting in estimates of the ionization energies of Mg$_i^0$ and Mg$_i^+$ of $\SI{107.50}{meV}$ and $\SI{256.49}{meV}$, respectively, in good agreement with those of Ho and Ramdas \cite{Ho1972}. They also observed a small splitting of the 3p$_{\pm}$ line, and ascribed it to a central cell splitting, as for 2p$_{\pm}$ \cite{Ho1993}. These authors also demonstrated that weak absorption lines observed in some of the earlier studies resulted from the neutral and singly ionized charge states of a double donor resulting from a Mg$_i$-O complex, with ionization energies of 124.66 and 274.90 meV, respectively \cite{Ho1998,Ho2003}. The formation of such complexes is a familiar theme for rapidly diffusing interstitial species. The Mg$_i$-O double donor for instance has similarities to the Li$_i$-O shallow donor which, as for the Mg$_i$ case, has a larger ionization energy than does the isolated interstitial Li shallow donor \cite{Pajot2010}. \par

A number of previously reported, relatively weak donor-like absorption transitions lying just below the dominant Mg$_i^0$ and  Mg$_i^+$ features in energy are a particular focus of this study. Lin \cite{Lin1982} first reported a series of features interpreted as the 2p$_0$, 2p$_{\pm}$, 3p$_0$, 3p$_{\pm}$ and 4p$_{\pm}$ transitions of a neutral donor they labelled $X_4$, with ionization energy $\sim\SI{93}{meV}$. Ho \cite{Ho2003,Ho2006} further studied this center, labelling the 2p$_0$, 2p$_{\pm}$, and 3p$_{\pm}$ transitions as lines $``\textrm{1}"$, $``\textrm{3}"$ and $``\textrm{5}"$ and determining an ionization energy of $\SI{93.57}{meV}$. Ho \cite{Ho2003,Ho2006} also reported the 2p$_0$, 2p$_{\pm}$, and 3p$_{\pm}$ of what was interpreted as a different neutral donor with an ionization energy $\SI{94.36}{meV}$. These were labelled lines $``\textrm{2}"$, $``\textrm{4}"$, and $``\textrm{6}"$. Ho \cite{Ho2006} further reported a doublet absorption feature labelled line $``\textrm{a}"$, and a singlet labelled line $``\textrm{b}"$, which were interpreted as the 2p$_{\pm}$ and 3p$_{\pm}$ transitions of an unknown singly-ionized double donor with ionization energy $\SI{213.53}{meV}$. \par

We will retain the $``\textrm{1}"$ through $``\textrm{6}"$ and $``\textrm{a}"$ and $``\textrm{b}"$ labels for these transitions, but will instead show that they all arise from the neutral and singly ionized states of a single perturbed Mg$_i$ donor with symmetry reduced below T$_d$. This could be a consequence of either Mg$_i$ inhabiting an alternate interstitial site, as detailed by Baxter and Ascarelli, \cite{Baxter1973} or Mg$_i$ complexing with other species. Evidence detailed later points to a complex as the more likely scenario. We denote these centers as Mg$_{i*}^0$ (including lines $``\textrm{1}"$-$``\textrm{6}"$) and Mg$_{i*}^+$ (including lines $``\textrm{a}"$ and $``\textrm{b}"$) for the neutral and singly-ionized charged states of the double donor. Mg$_{i*}^0$ is unusual in that even though the ground state binding energy is larger that that of normal shallow donors, it has very low-lying excited state components which can be thermally populated at relatively low temperatures. This gives rise to lines $``\textrm{1}"$, $``\textrm{3}"$, and $``\textrm{5}"$. For normal shallow donors having T$_d$ symmetry, the 1s(A$_1$) singlet ground state is much further separated from the 1s(E) and 1s(T$_2$) valley-orbit excited states. \par

Given the large number of as yet unidentified absorption transitions reported in Mg-diffused Si, one of the goals of this study was to identify species which were intrinsic to Mg$_i$, and not complexes with other impurities.  This was achieved by using well-characterized, high purity, float-zone Si starting material, and high purity Mg as the diffusion source.  As a result, absorption from Mg-O complexes \cite{Ho1998,Ho2003} was either unobservable or very weak. A second goal was to investigate the possible spectral linewidth improvements in isotopically enriched 28-silicon ($^{28}$Si), which, by eliminating inhomogeneous isotope broadening, produces remarkable improvements in linewidth for some transitions of both shallow donors and deep chalcogen donors \cite{Morse2017,Karaiskaj2003,Cardona2005,Steger2011,Steger2009}. The final goal was to search for the as yet undetected 1s(A$_1$) to 1s(T$_2$) absorption transition, which is forbidden under the effective mass approximation, but symmetry allowed, and very strong for the deep chalcogen double donors. \par

No evidence for the 1s(A$_1$) to 1s(T$_2$) absorption transition could be observed for either Mg$_i^0$ or Mg$_i^+$, but the expected location for these transitions suffers from overlap with other relatively strong spectral features. Attempts to observe the 1s(A$_1$) to 1s(E) transition for either charge state of Mg$_i$ using electronic Raman scattering were also unsuccessful. The Mg$_{i*}$ counterparts of Mg$_i$ likewise revealed no sign of the s-like excited states. Some spectral lines revealed small energy shifts and linewidth improvements in $^{28}$Si, although not as dramatic as that seen for shallow impurities or the deep chalcogen double donors. \par 

Even though our goal was to investigate transitions intrinsic to Mg$_i$ impurities, the acceptor boron (B) is an omnipresent impurity in silicon, and higher B concentrations were intentionally used to create higher concentrations of Mg$_i^+$. This resulted in the discovery of absorption related to a new shallow donor impurity which we ascribe to a interstitial magnesium-substitutional boron pair. This new Mg$_i$-B$_s$ shallow donor can bind excitons and is observed in the photoluminescence spectra of these samples, where the donor binding energy and exciton localization energy are found to follow Hayne’s Rule \cite{Haynes1960} for Si. \par

\FloatBarrier
\section{Materials and methods}
Shuman et al. have published recent results \cite{1Shuman2017,2Shuman2017} detailing parameters for Si:Mg diffusion. In this study we worked with four float-zone grown diffused Si:Mg samples. These included two relatively high boron content $^{28}$Si and $^{\textrm{nat}}$Si samples with boron concentrations of 1.8$\times$10$^{15}$ cm$^{-3}$ and 2.2$\times$10$^{15}$ cm$^{-3}$ respectively ($^{28}$Si HB and $^{\textrm{nat}}$Si HB), and two undoped low boron content $^{28}$Si and $^{\textrm{nat}}$Si samples ($^{28}$Si LB and $^{\textrm{nat}}$Si LB) with boron concentrations of approximately 4$\times$10$^{13}$ cm$^{-3}$ and 1$\times$10$^{13}$ cm$^{-3}$ respectively. The properties of the $^{28}$Si LB material have been detailed elsewhere \cite{Devyatykh2008}. Both $^{28}$Si samples were enriched to $\SI{99.995}{\%}$ $^{28}$Si. \par

All absorption and photoluminesence measurements were performed using a Bruker IFS 125HR Fourier transform infrared (FTIR) spectrometer. Samples were mounted in a liquid helium cryostat with either polypropylene or ZnSe windows. Pumped helium sub-lambda temperatures were used for all absorption and luminescence measurements, except where noted. For infrared absorption spectra, a KBr or coated Mylar beam-splitter was used, with a liquid nitrogen-cooled mercury cadmium telluride detector when studying Mg$_i^0$/Mg$_i^+$ centers, and a 4.2 K silicon bolometer with an 800 cm$^{-1}$ low-pass cold filter for studying the shallow donor region. \par

\FloatBarrier
\section{Experimental results and discussion}
We present our results in the following order: First, we discuss our spectra displaying the well known Mg$_{i}^0$/Mg$_{i}^+$ donor transitions previously noted in the literature and show a number of higher excited states for each. Second, we present our study of the Mg$_{i*}^0$/Mg$_{i*}^+$ transitions first documented by Ho \cite{2Ho2003,Ho2006} which we interpret to most likely represent Mg complexing with some other species resulting in a symmetry less than T$_d$. Here we also note the presence of yet another set of lines also previously seen in Ho \cite{2Ho2003,Ho2006} which appear to be thermally induced transitions from Mg$_{i*}^0$. The labelling scheme adopted by Ho \cite{2Ho2003,Ho2006} for lines $``\textrm{1}"$-$``\textrm{6}"$ and $``\textrm{a}"$, $\textrm{``b"}$ is maintained in all figures where those transitions are visible together with our own identifications. Third, we detail our observations of a new shallow donor species seen in absorption and photoluminesence which we interpret as a magnesium-boron complex. \par

In each figure we choose spectra that best display the features relevant to each plot, i.e. successive plots are not necessarily different segments of the same spectra. Peak positions listed in Tab.~\ref{tab:Mg0Donors} and \ref{tab:Mgpl=Donors} are obtained from fits to spectra that best show the specific feature in question. All peak energies determined from fits are assumed to carry an uncertainty of $\SI{\pm 0.01}{meV}$ unless otherwise specified. For all figures, spectra are presented from top to bottom in the order specified in the legend. Scaling factors associated with a given spectrum indicate that a trace has been multiplied by the listed value. While the spectra have been corrected for instrumental throughput, they are not corrected for reflection and scattering losses, or for multiphonon absorption. The spectra are plotted vs. relative absorption coefficient, where all that matters is the height of lines above their baselines. Offsets are applied for ease of viewing in most spectra where multiple traces are compared. Unless otherwise specified, spectra were collected at $T$ = 2.1 K with 0.1 cm$^{-1}$ ($\sim$ 0.012 meV) resolution. \par

Illumination conditions on samples were variable since our apparatus did not keep our samples in the dark and relies on broad-band IR illumination to perform spectroscopy. We were not able to reproduce the results of Baxter and Ascarelli, \cite{Baxter1973} which suggested that we should be able to photoconvert T$_d$ magnesium interstitials to C$_{3v}$ with sufficient power from an IR laser. We attempted this with up to $\SI{30}{mW}$ of power, supplied with an IPG photonics IR laser at $\SI{2.2}{\mu m}$. \par

\FloatBarrier
\subsection{Mg$_i^{0}$ and Mg$_i^{+}$ donor spectra}
Here we display absorption data showing lines previously seen \cite{Franks1967,Ho1972,Ho1993,Ho1998,Ho2003,2Ho2003,Ho2006} for the standard Mg$_i$ center and a number of newly resolved higher excited states. We observe no sign of Mg$_i$ 1s(A$_1$) to 1s(T$_2$) transitions for the neutral/ionized species. Calculations by Altarelli \cite{Altarelli1983} of binding energies for excited states of singly-ionized double donors in silicon suggest a binding energy of 155 meV for 1s(T$_2$). For the case of Mg$_i^+$ this would lead us to anticipate 1s(T$_2$) appearing in the Mg$_i^0$ region. Magnesium, while a relatively deep donor, is not as deep as chalcogens like, Se$^+$ (593 meV) \cite{Morse2017,Steger2009}, for which these transitions are very strong. As a result, transitions within the `s' states are likely better approximated by EMT and so more strongly forbidden. \par 

\begin{figure}[htbp!]
\includegraphics[width=0.42\textwidth]{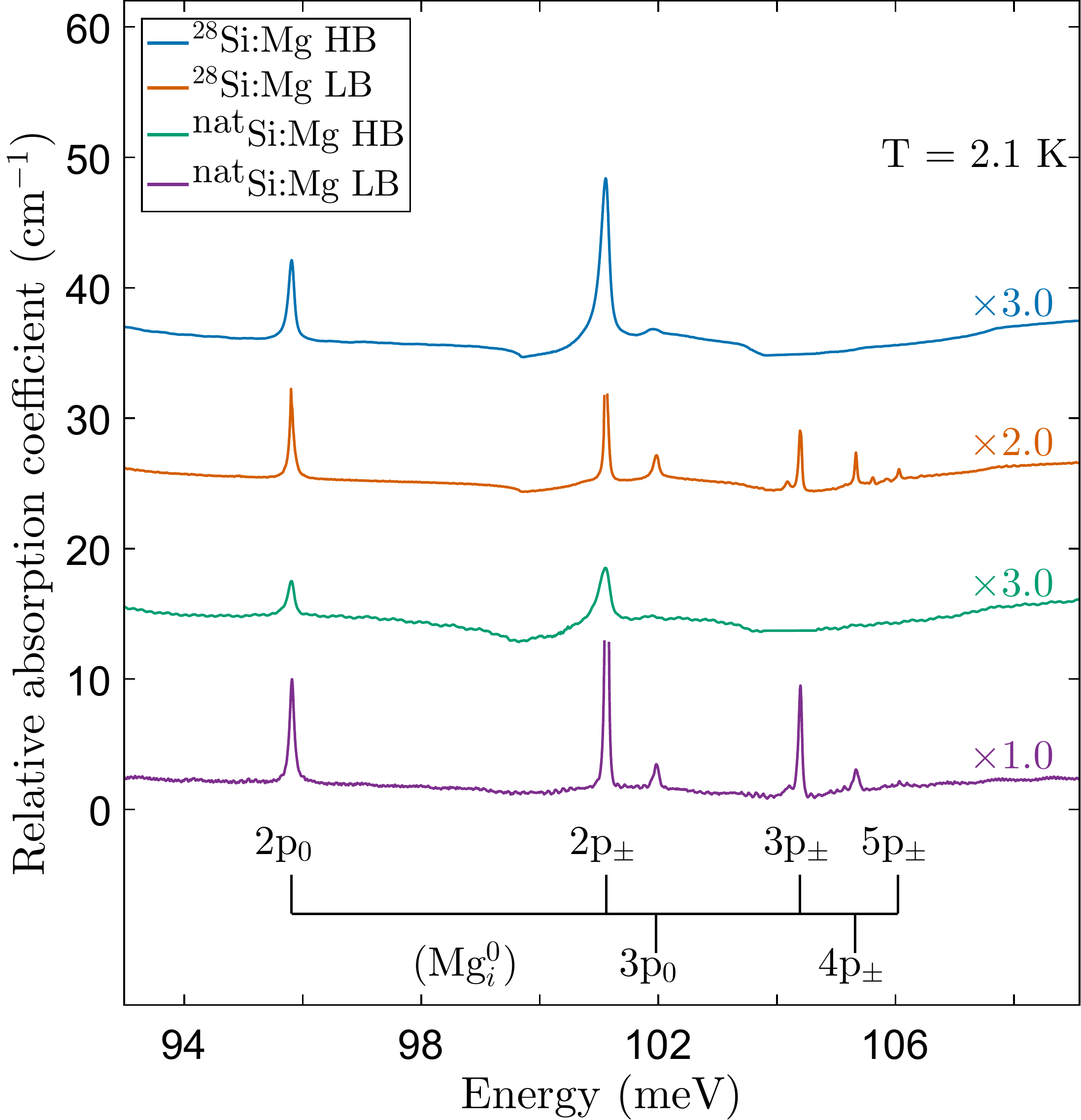}
\caption{Here we examine the Mg$_i^0$ region. The 2p$_{\pm}$ transitions have been truncated in LB spectra since for our sample thicknesses, the transmission there was essentially zero. Excited states are visible as high as 6p$_{\pm}$ (See Fig.~\ref{highExc} for a more detailed view) in LB samples, but no higher than 3p$_{0}$ in HB samples due to Stark broadening.}
\label{Mg0zoom}
\end{figure}

The Mg$_i^0$ and Mg$_i^+$ regions reveal many excited states in LB samples, with HB counterparts characterized by Stark broadened features and a resulting absence of higher excited states. This is seen in Fig.~\ref{Mg0zoom} and further detailed in Tab.~\ref{tab:Mg0Donors} for Mg$_i^0$, which for HB samples have broad features with no peaks beyond 2p$_{\pm}$. \par

\begin{figure}[htbp!]
\includegraphics[width=0.42\textwidth]{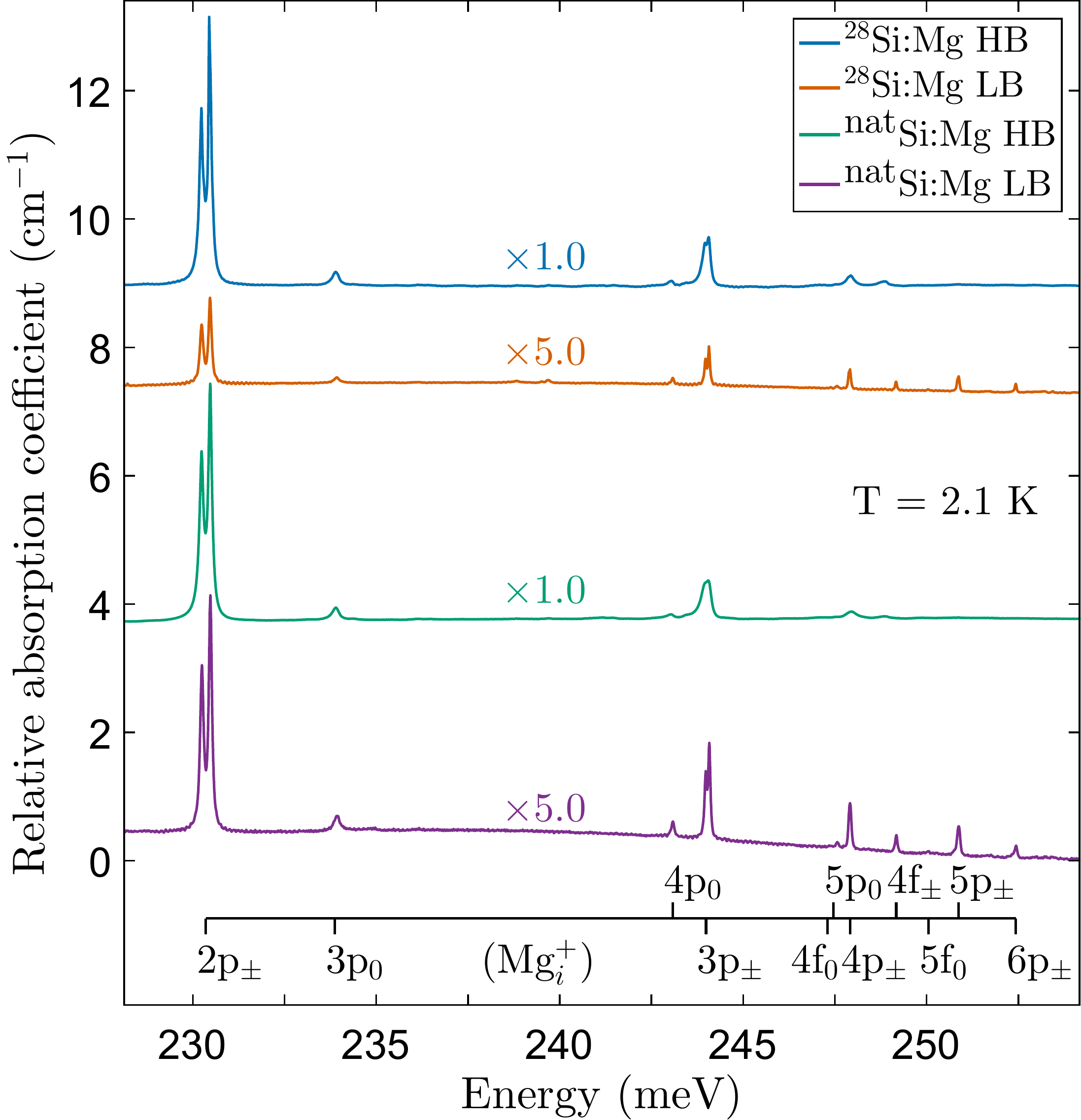}
\caption{A closer look at the transitions of Mg$_i^+$ reveals p$_{\pm}$ doublets (most obviously for 2p$_{\pm}$) suggested previously to result from central cell splitting. The splitting decreases steadily through higher excited states, and is less obvious in HB samples due to Stark broadening.}
\label{Mg+zoom}
\end{figure}

Similarly in the Mg$_i^{+}$ region seen in Fig.~\ref{Mg+zoom} we note many higher excited states, some newly observed, which are obscured by Stark broadening in more heavily compensated samples. Of particular note in these spectra is splitting of the p$_{\pm}$ levels visible for 2p$_{\pm}$ in all samples and in higher p$_{\pm}$ levels in LB samples. Our $^{28}$Si:Mg LB spectra have afforded us an opportunity to study a number of very weak transitions for Mg$_i^+$, detailed in Tab.~\ref{tab:Mgpl=Donors}, that have not been seen previously. With its lack of isotopic broadening and relatively low boron compensation, this $^{28}$Si sample has also allowed us to observe splitting of the p$_{\pm}$ levels as high as 6p$_{\pm}$. As seen in Tab.~\ref{tab:Mgpl=Donors}, these splittings are 0.23, 0.10, 0.04, 0.04, and $\SI{0.02}{ meV}$ for 2p$_{\pm}$, 3p$_{\pm}$, 4p$_{\pm}$, 5p$_{\pm}$, and 6p$_{\pm}$ respectively. In previous Mg$_i$ studies doublets have been proposed to result from central cell splitting of these states \cite{Ho1972,Thilderkvist1994}. Our observations are consistent with decreasing central cell overlap and correspondingly decreased splitting with higher excited state. As determined by Kohn and Luttinger \cite{Kohn1955}, in the T$_d$ symmetry group the 2p$_{\pm}$ level has the reducible representation 2T$_1$+2T$_2$. From the ground state, only transitions to states with T$_2$ symmetry are allowed, and lifting of degeneracy between the two T$_2$ levels through central cell interaction leads to the commonly observed doublet seen in Fig.~\ref{Mg+zoom} for 2p$_{\pm}$ and, to a lesser degree, for higher p$_{\pm}$ levels. \par

\begin{figure}[htbp!]
\includegraphics[width=0.44\textwidth]{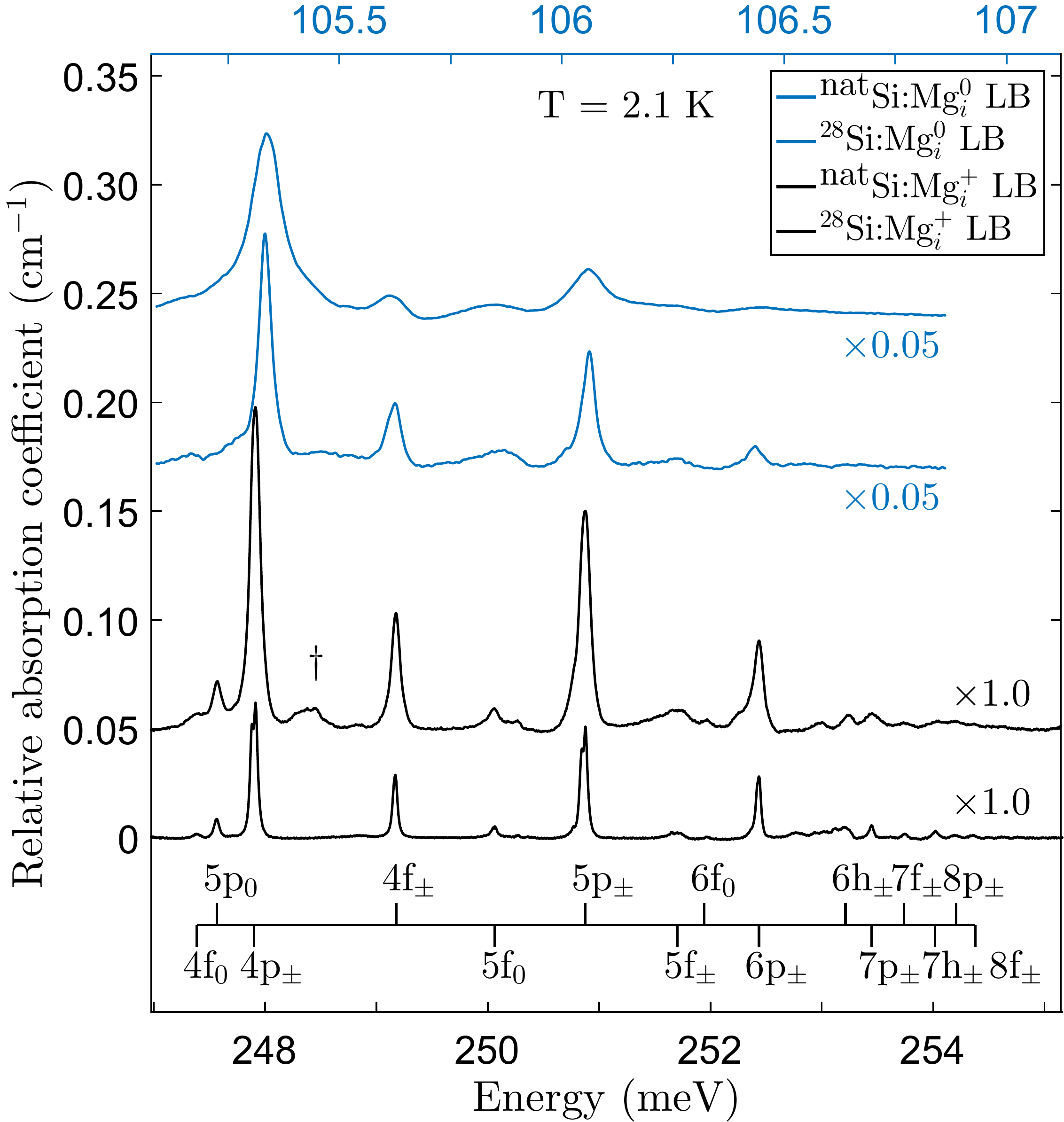}
\caption{Spectra displaying several newly observed excited states of Mg$_i^{0}$ and Mg$_i^{+}$ in LB samples. The bottom axis (black) corresponds to Mg$_i^{+}$ spectra while top (blue) corresponds to Mg$_i^{0}$. The spectral window for the Mg$_i^{+}$ data displayed has been chosen to be four times larger than that of Mg$_i^{0}$, giving matched line spacings for ease of viewing. The $\dagger$ symbol labels a feature of unknown origin seen in the $^{\textrm{nat}}$Si Mg$_i^{+}$ spectrum. Spectra were collected at 0.05 cm$^{-1}$ ($\sim$ 0.0062 meV) resolution.}
\label{highExc}
\end{figure}

\begin{figure}[htbp!]
\includegraphics[width=0.42\textwidth]{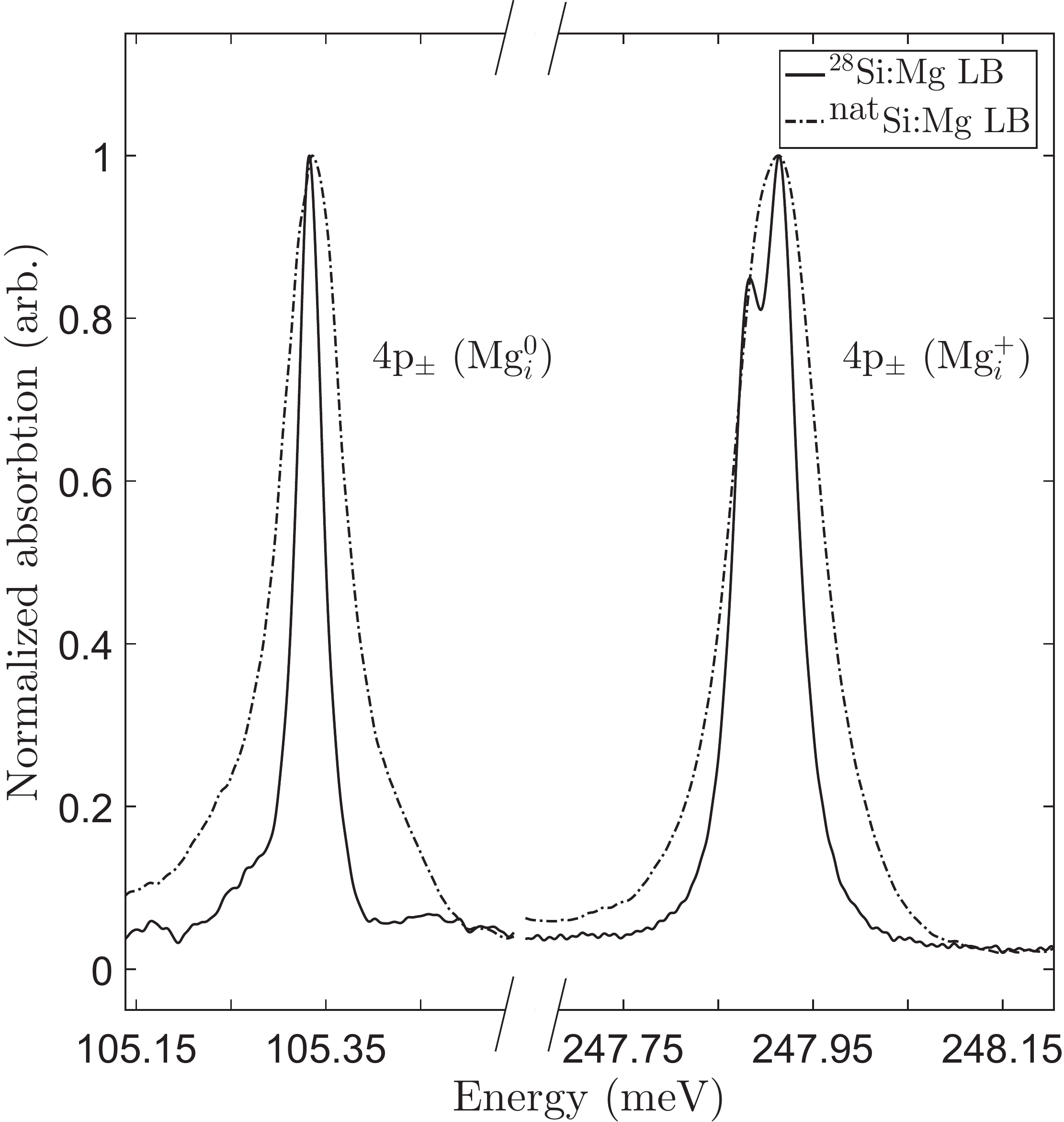}
\caption{Here we illustrate the narrower linewidths observed in $^{28}$Si due to lack of isotopic broadening, and a shift in ground state energy of $\sim$ 0.01 meV between $^{28}$Si and $^{\textrm{nat}}$Si. This is shown above for the 4p$_{\pm}$ states of Mg$_i^{0}$ and Mg$_i^{+}$. The Mg$_i^{0}$ and Mg$_i^{+}$ are presented on a split axis with all lines normalized to the same height for ease of comparison.}
\label{isotopeShift}
\end{figure}

These splittings of p$_{\pm}$ states can be seen in Figs.~\ref{Mg+zoom}, \ref{highExc}, and \ref{isotopeShift}. The energies for these and all other labelled transitions are included in Tabs.~\ref{tab:Mg0Donors} and \ref{tab:Mgpl=Donors} for neutral and ionized donors respectively. Transitions in the ionized species are in general less impacted by Stark broadening in HB samples relative to counterparts in the neutral species, Mg$_i^0$, due to lower electric-field sensitivity in the more tightly bound singly-ionized donor excited states. The heightened electric field sensitivity in the neutral species results in no observable splitting of the Mg$_i^0$ p$_{\pm}$ states.

\begin{figure}[htbp!]
\includegraphics[width=0.42\textwidth]{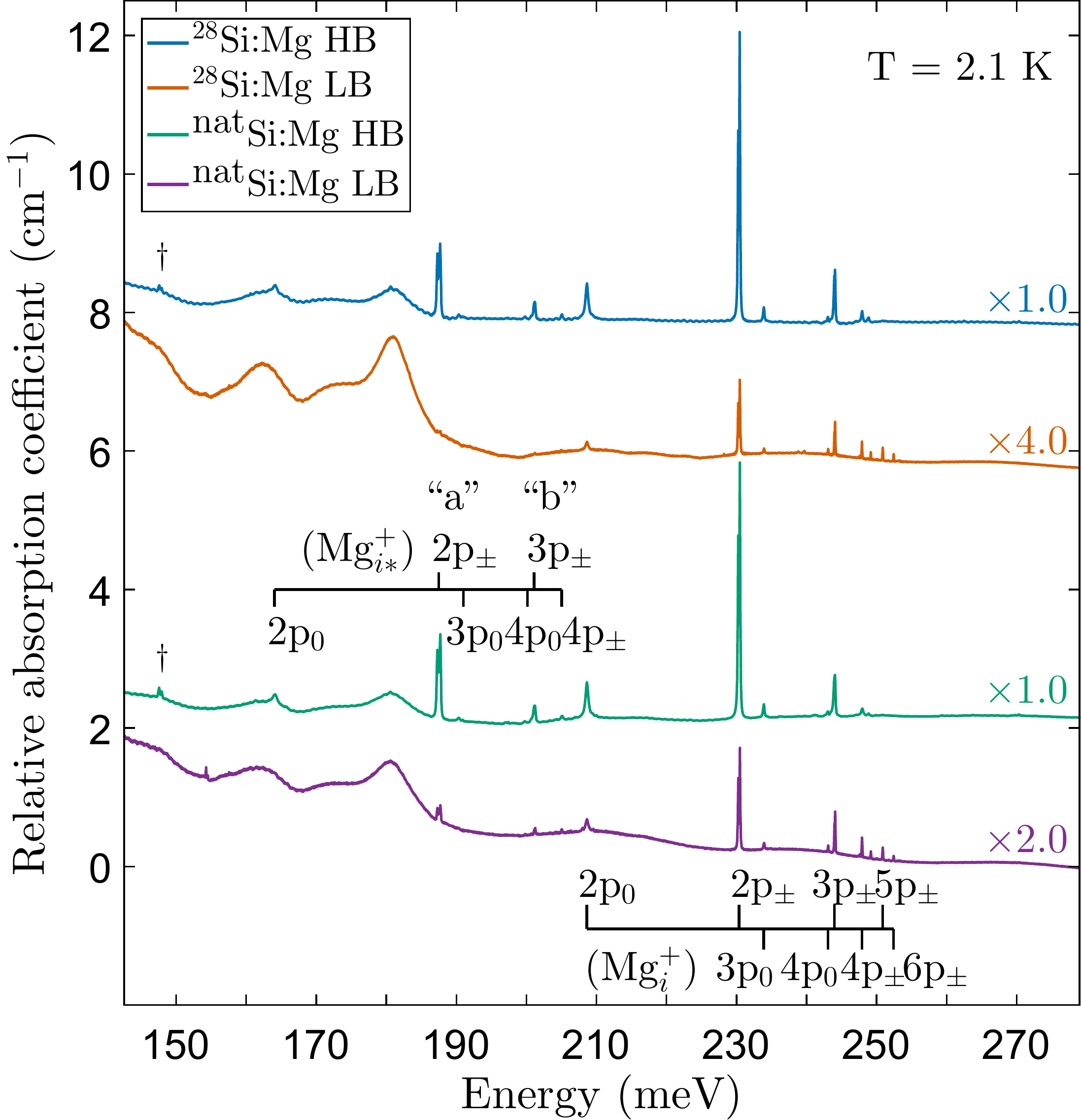}
\caption{In the ionized region we observe known excited states as high as 6p$_{\pm}$ for Mg$_i^{+}$ and up to 4p$_{\pm}$ for the perturbed magnesium center, Mg$_{i*}^{+}$, which is examined more closely later. The 2p$_{\pm}$ and 3p$_{\pm}$ levels of Mg$_{i*}^{+}$ are also labelled by $``\textrm{a}"$ and $``\textrm{b}"$ for consistency with the labelling scheme introduced by Ho \cite{Ho2006}. A doublet of unknown origin is visible and indicated by a $\dagger$ in both $^{\textrm{nat}}$Si and $^{28}$Si HB samples.}
\label{Mg+star}
\end{figure}

In Fig.~\ref{Mg+star} we see a broader view of the Mg$_i^{+}$ spectrum, including 2p$_0$ and also note the presence of another singly ionized double donor labelled Mg$_{i*}^{+}$, the data for which is covered in the next section. We see an doublet with peaks at 147.59 meV and 147.95 meV, indicated in Fig.~\ref{Mg+star} with a $\dagger$ in HB samples. This is the weakly visible 2p$_{\pm}$ of an unidentified donor.

In Fig.~\ref{isotopeShift} we note the small shift between transitions in $^{28}$Si and $^{\textrm{nat}}$Si generated by an $\sim$ 0.01 meV difference in their ground state binding energy. A slight narrowing of the peaks in isotopically enriched silicon is visible relative to its natural silicon counterparts for both Mg$_i^{0}$ and Mg$_i^{+}$. We also note that even in LB samples, neither natural nor enriched silicon samples show particularly narrow transitions for the singly-ionized species relative to those of the singly-ionized chalcogen double donors \cite{Morse2017,Steger2011,Steger2009}. This may be a consequence of magnesium also incorporating as a substitutional double acceptor \cite{Baber1988} as well as an interstitial double donor, leading to additional strain fields and Stark broadening even in LB samples. \par

\FloatBarrier
\subsection{Mg$_{i*}^{0}$ and Mg$_{i*}^{+}$ donor spectra}
\begin{figure}[htbp!]
\includegraphics[width=0.42\textwidth]{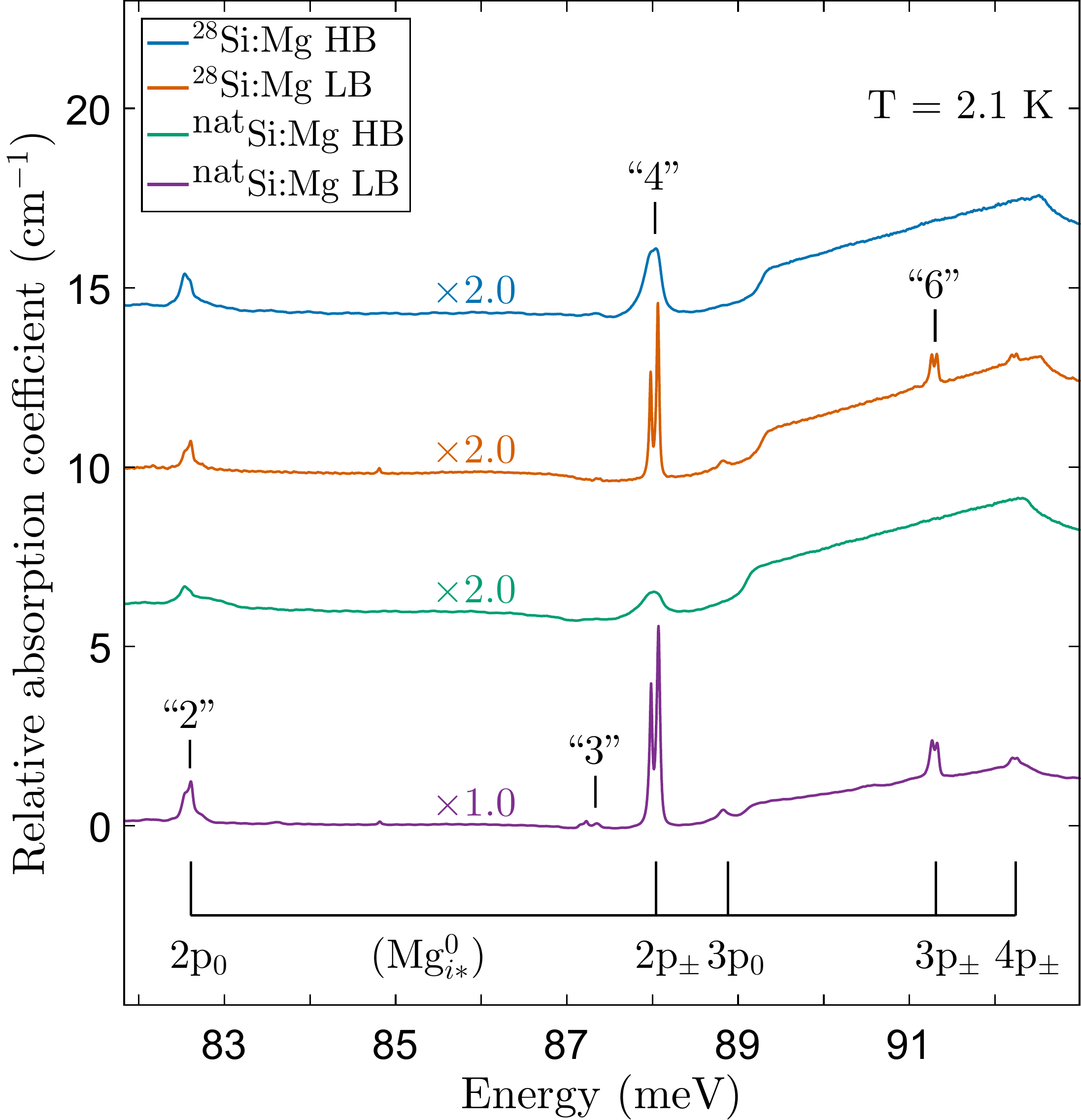}
\caption{The Mg$_{i*}^0$ center (with excited states seen up to 4p$_{\pm}$) has been previously observed by Ho \cite{2Ho2003,Ho2006}. We note that the 2p$_0$ transition has partially resolved fine structure suggesting up to four peaks as detailed in Fig.~\ref{Mg0st_fineStruct}. The labels $``\textrm{2}"$, $``\textrm{3}"$, $``\textrm{4}"$, and $``\textrm{6}"$ are those introduced by Ho \cite{2Ho2003,Ho2006}.}
\label{Mg0star}
\end{figure}

Here we include views of the neutral and singly ionized species of the double donor, Mg$_{i*}$, which most likely represents magnesium complexing with another impurity. The resulting reduction in symmetry relative to T$_d$ may explain the additional fine structure we observe in a number of spectral lines. This section also contains tables of transition energies as seen in natural and isotopically enriched Si samples for all Mg$_{i}^{0}$ and Mg$_{i*}^{0}$ lines (Tab.~\ref{tab:Mg0Donors}) as well as Mg$_{i}^{+}$ and Mg$_{i*}^{+}$ (Tab.~\ref{tab:Mgpl=Donors}). Background structure seen in the Mg$_i$/Mg$_{i*}$ regions is generated by a broad band of multi-phonon absorption features seen most prominently in Figs.~\ref{Mg+star} and \ref{Mg0star}.

\begin{figure}[htbp!]
\includegraphics[width=0.42\textwidth]{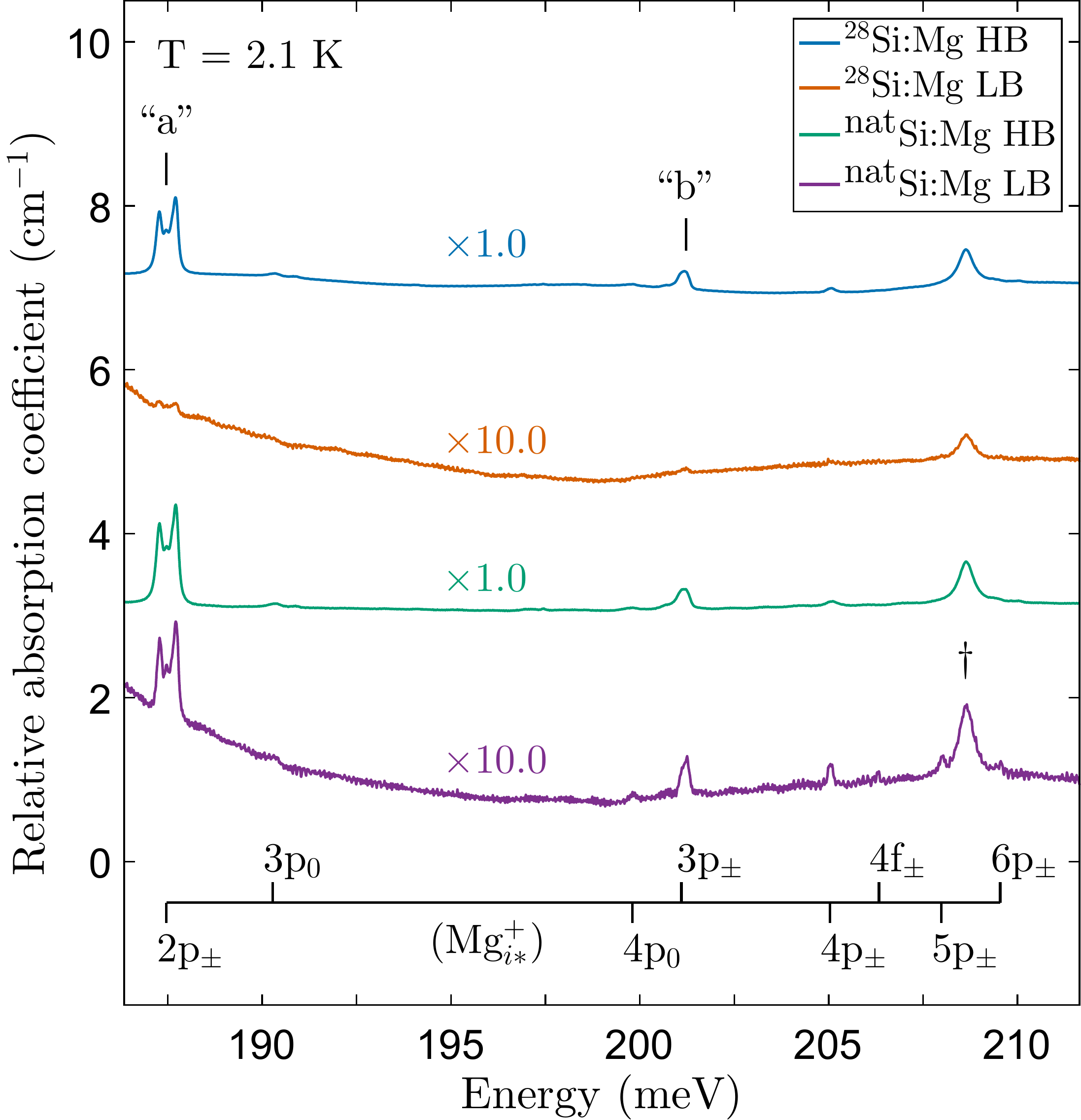}
\caption{Here we see the absorption spectrum of Mg$_{i*}^{+}$. Note that there is more than a doublet structure seen in the 2p$_{\pm}$ line of Mg$_{i*}^+$. Closer inspection reveals four peaks, as seen in Fig.~\ref{Mg+st_fineStruct}. The $\dagger$ symbol denotes the 2p$_0$ line of Mg$_{i}^{+}$. Labels $``\textrm{a}"$ and $``\textrm{b}"$ correspond to Ho's notation \cite{Ho2006} for lines we identify as the 2p$_{\pm}$ and 3p$_{\pm}$ of Mg$_{i*}^{+}$.}
\label{Mg+stzoom}
\end{figure}

Below the transitions of Mg$_{i}^{0}$ and Mg$_{i}^{+}$ seen in the previous section we observe the neutral and singly ionized species of another double donor, which we label Mg$_{i*}^{0}$ and Mg$_{i*}^{+}$. The neutral region is seen in Fig.~\ref{Mg0star} and the ionized in Figs.~\ref{Mg+star} and \ref{Mg+stzoom}. Some of these lines have been observed previously by Ho \cite{2Ho2003,Ho2006} who labelled features of our Mg$_{i*}^{0}$ as lines $``\textrm{1}"$-$``\textrm{6}"$, and of Mg$_{i*}^{+}$ as $``\textrm{a}"$ and $``\textrm{b}"$, however our interpretation of all these lines as arising from the neutral and singly ionized charge states of a single perturbed Mg$_i$ species is new.

As for Mg$_{i}^{0}$ and Mg$_{i}^{+}$, the HB samples show significant Stark broadening that obscures any fine structure, and suppresses higher excited states of the Mg$_{i*}$ species. High excited states are most easily resolved in the LB sample spectra. \par

We determine the ionization energies of Mg$_{i*}^0$ and Mg$_{i*}^+$ by the binding energies of one of the highest cleanly resolved Mg$_{i*}$ transitions, 4p$_{\pm}$, added to the corresponding EMT predicted binding energies. For 4p$_{\pm}$ these are 2.187 and 8.75 meV for the neutral and ionized species respectively \cite{Pajot2010}. From this we estimate that Mg$_{i*}^0$ and Mg$_{i*}^+$ have ionization energies of 94.41 and $\SI{213.80}{meV}$, both slightly higher than the estimates given by Ho, which were 94.36 and $\SI{213.53}{meV}$ \cite{Ho2003,Ho2006}.

\begin{figure}[htbp!]
\includegraphics[width=0.42\textwidth]{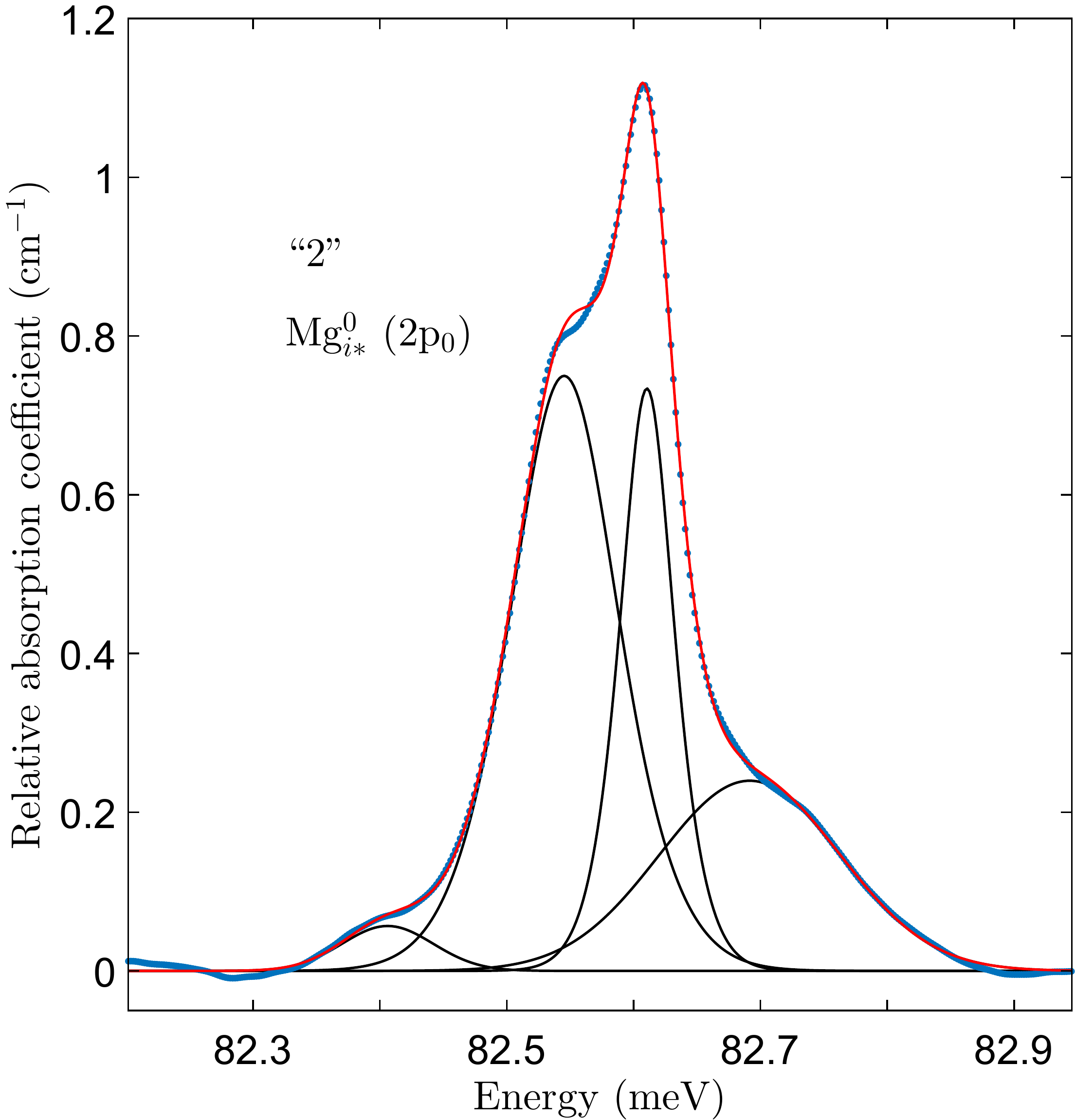}
\caption{A fit to the fine structure of the 2p$_{0}$ feature of Mg$_{i*}^0$, referred to by Ho \cite{Ho2006} as line $``2"$, using four mixed Gaussian-Lorentzian peaks. Data shown corresponds to the $^{\textrm{nat}}$Si LB sample.}
\label{Mg0st_fineStruct}
\end{figure}

We note unusual fine structure in the 2p$_{0}$ level of Mg$_{i*}^0$ seen in Fig.~\ref{Mg0star} and through fits to the LB sample 2p$_{0}$ transition in Fig.~\ref{Mg0st_fineStruct}. Energies of the peaks in this apparent partially resolved quartet are given in Tab.~\ref{tab:Mg0Donors}. In T$_d$ symmetry, the 2p$_{0}$ level has the representation A$_1$ + E + T$_2$. Of these however, only transitions from an A$_1$ ground state to states with T$_2$ symmetry are dipole allowed in the tetrahedral group. The fine structure we observe is therefore immediately indicative of a symmetry lower than T$_d$. \par

\begin{table}[htbp!]
\begin{ruledtabular}
\begin{tabular}{l c c c c}
& \multicolumn{4}{c}{Transition energy (meV)} \\
\cline{2-5}
Label & Mg$_i^0$ ($^{\textrm{nat}}$Si) & Mg$_{i*}^0$ ($^{\textrm{nat}}$Si) & Mg$_i^0$ ($^{28}$Si) & Mg$_{i*}^0$ ($^{28}$Si) \\
\hline
2p$_0$ & 95.82 & \makecell[t]{82.40 \\ 82.55 \\ 82.61 \\ 82.69} & 95.81 & \makecell[t]{ - \\ 82.55 \\ 82.61 \\ 82.71$\dagger$} \\
2p$_{\pm}$ & 101.08 & \makecell[t]{87.98 \\ 88.07} & 101.09  & \makecell[t]{87.98 \\ 88.06} \\
3p$_0$     & 101.97 & 88.82 & 101.96 & 88.82 \\
4p$_0$     & 104.24 & - & 104.18 & - \\
3p$_{\pm}$ & 104.39 & \makecell[t]{91.27 \\ 91.33} & 104.40 & \makecell[t]{91.26 \\ 91.32} \\
4f$_{0}$ & - & - & - & - \\
5p$_{0}$ & - & - & - & - \\
4p$_{\pm}$ & 105.34 & \makecell[t]{92.20 \\ 92.26} & 105.33 & \makecell[t]{92.19 \\ 92.25} \\
4f$_{\pm}$ & 105.61 & - & 105.62 & - \\
5f$_{0}$ & 105.84 & - & 105.85 & - \\
5p$_{\pm}$ & 106.06 & 92.95$\dagger$ & 106.06 & - \\
5f$_{\pm}$ & 106.22 & - & 106.25 & - \\
6f$_{0}$ & - & - & - & - \\
6p$_{\pm}$ & 106.47$\dagger$ & - & 106.43$\dagger$ & - \\
\end{tabular}
\end{ruledtabular}
\caption{Peak positions of observed Mg$_i^{0}$/Mg$_{i*}^{0}$ transitions in $^{\textrm{nat}}$Si and $^{28}$Si samples. The $\dagger$ symbol is used to label particularly weak transitions.}
\label{tab:Mg0Donors}
\end{table}

\begin{figure}[htbp!]
\includegraphics[width=0.42\textwidth]{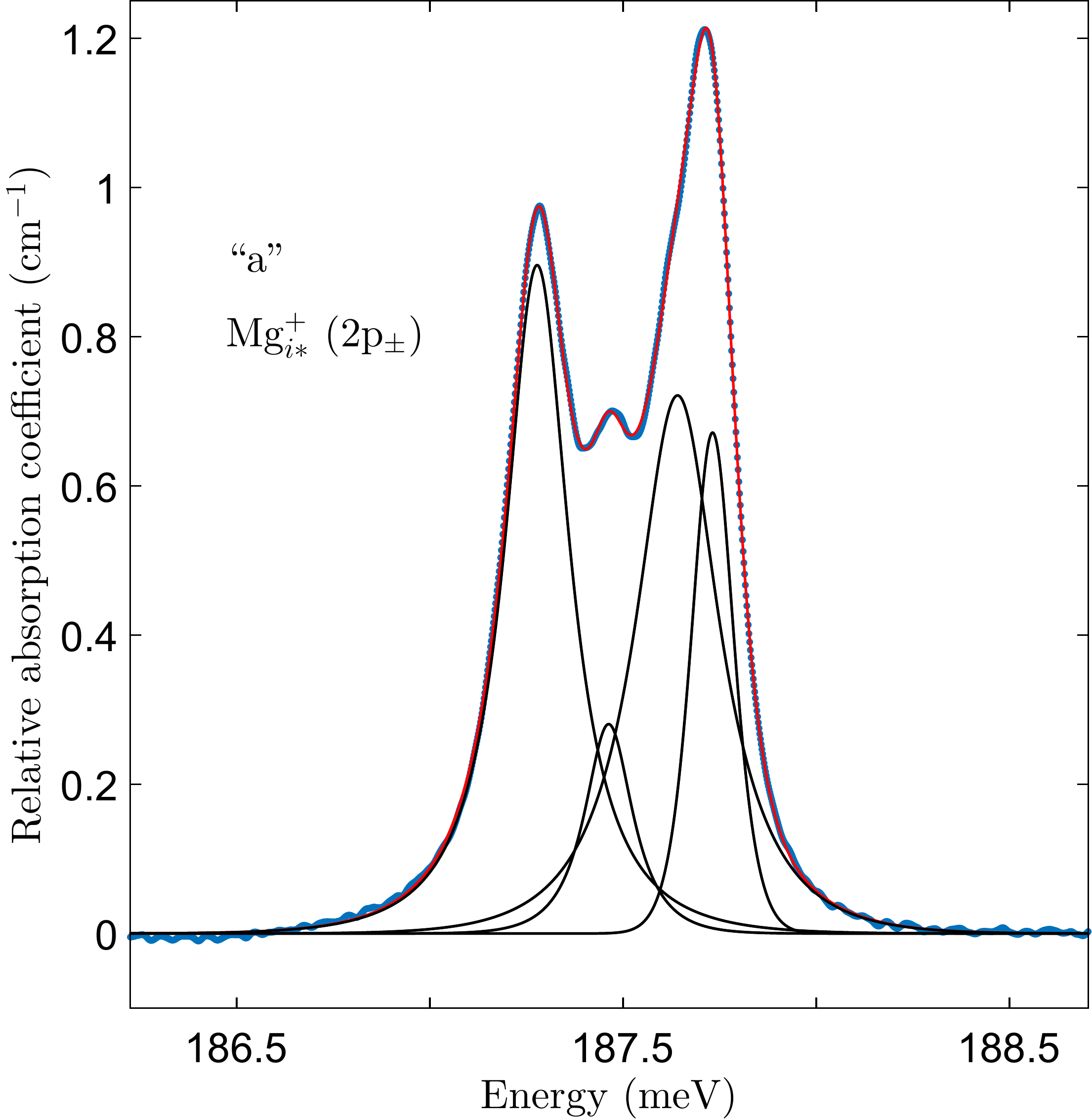}
\caption{A fit to the fine structure of the 2p$_{\pm}$ feature of Mg$_{i*}^+$, referred to by Ho \cite{Ho2006} as line $``\textrm{a}"$, using four mixed Gaussian-Lorentzian peaks. Data shown corresponds to the $^{\textrm{nat}}$Si HB sample.}
\label{Mg+st_fineStruct}
\end{figure}

As with the standard Mg$_{i}^0$/Mg$_{i}^+$ species, Mg$_{i*}^+$ is significantly less sensitive to Stark broadening than Mg$_{i*}^0$. The absence of this broadening in Mg$_{i*}^+$, combined with central cell interactions that lift degeneracy between states allowed by the new low symmetry configuration \cite{Thilderkvist1994}, can lead to complex fine structure. This is visible in the 2p$_{\pm}$ of Mg$_{i*}^{+}$ seen in Fig.~\ref{Mg+stzoom}, which appears as a partially resolved quartet of states rather than the doublet seen in Mg$_{i}^{+}$. A fit to this structure is included in Fig.~\ref{Mg+st_fineStruct}. The Mg$_{i*}^+$ 2p$_{0}$ feature was rather weak and broad as seen in Fig.~\ref{Mg+star}, and no fine structure could be resolved.\par

We note that the observation of doublets for the p$_{\pm}$ levels of Mg$_{i*}^{0}$ is not necessarily inconsistent with the quartet structure of the 2p$_{\pm}$ level of Mg$_{i*}^{+}$. Splittings are smaller due to reduced central cell interaction and Stark broadenings are greater due to increased electric field sensitivity. More generally, the observed splittings for all of these lines can only put a lower bound on the number of non-degenerate components of a given transition. \par

Splitting of the p$_{\pm}$ levels for Mg$_{i*}^0$ spectra is visible only in LB samples, with both $^{\textrm{nat}}$Si/$^{28}$Si spectra revealing doublets for several p$_{\pm}$ levels. Peak values for visible splittings as high as 4p$_{\pm}$ in Mg$_{i*}^{0}$ are detailed in Tab.~\ref{tab:Mg0Donors}, with splittings of 0.09, 0.06, and $\SI{0.06}{meV}$ for 2p$_{\pm}$, 3p$_{\pm}$, and 4p$_{\pm}$ respectively as seen in Fig.~\ref{Mg0star}. In Fig.~\ref{Mg0star}, we include Ho's labels $``\textrm{2}"$, $``\textrm{4}"$, and $``\textrm{6}"$ \cite{2Ho2003,Ho2006} corresponding to the 2p$_0$, 2p$_{\pm}$, and 3p$_{\pm}$ of Mg$_{i*}^{0}$. We further indicate line $``\textrm{3}"$ also observed by Ho, which will be discussed in the following section. \par

\begin{table}[htbp!]
\begin{ruledtabular}
\begin{tabular}{l c c c c}
& \multicolumn{4}{c}{Transition energy (meV)} \\
\cline{2-5}
Label & Mg$_i^+$ ($^{\textrm{nat}}$Si) & Mg$_{i*}^+$ ($^{\textrm{nat}}$Si) & Mg$_i^+$ ($^{28}$Si) & Mg$_{i*}^+$ ($^{28}$Si) \\
\hline
2p$_0$ & 208.65 & 164.09 & 208.64 & 164.09 \\
2p$_{\pm}$ & \makecell[t]{230.25 \\ 230.48} & \makecell[t]{187.28 \\ 187.46 \\ 187.64 \\ 187.73} & \makecell[t]{230.24 \\ 230.47}  & \makecell[t]{187.27 \\ 187.46 \\ 187.63 \\ 187.73} \\
3p$_0$ & 233.94 & 190.33 & 233.92 & 190.32 \\
4p$_0$     & 243.08 & 199.76 & 243.08 & 199.77 \\
3p$_{\pm}$ & \makecell[t]{243.98 \\ 244.08} & 201.20 & \makecell[t]{243.97 \\ 244.07}  & 201.15 \\
4f$_{0}$ & 247.37 & - & 247.38 & - \\
5p$_{0}$ & 247.57 & - & 247.57 & - \\
4p$_{\pm}$ & 247.91 & 205.05 & \makecell[t]{247.88 \\ 247.92} & 205.05 \\
4f$_{\pm}$ & 249.17 & 206.30$\dagger$ & 249.17 & - \\
5f$_{0}$ & 250.05 & - & 250.06 & - \\
5p$_{\pm}$ & 250.87 & 208.00$\dagger$ & \makecell[t]{250.84 \\ 250.88} & - \\
5f$_{\pm}$ & 251.69 & - & 251.68 & - \\
6f$_{0}$ & 251.98 & - & 251.97 & - \\
6p$_{\pm}$ & 252.44 & 209.54$\dagger$ & \makecell[t]{252.41 \\ 252.43} & - \\
7f$_{0}$ & 252.98 & - & 253.02 & - \\
6h$_{\pm}$ & 253.23 & - & 253.21 & - \\
7p$_{\pm}$ & 253.45 & - & 253.44 & - \\
7f$_{\pm}$ & 253.73 & - & 253.74 & - \\
7h$_{\pm}$ & 254.04 & - & 254.02 & - \\
8p$_{\pm}$ & 254.19 & - & 254.20 & - \\
8f$_{\pm}$ & 254.37 & - & 254.35 & - \\
\end{tabular}
\end{ruledtabular}
\caption{Peak positions of observed Mg$_i^+$/Mg$_{i*}^+$ transitions in $^{\textrm{nat}}$Si and $^{28}$Si samples. The $\dagger$ labels particularly weak transitions.}
\label{tab:Mgpl=Donors}
\end{table}

The complex fine structure of Mg$_{i*}$ can be understood in the context of a reduced symmetry configuration. We speculate that these alternate centers result from Mg$_i$ forming a complex with another impurity. This new symmetry, that is a subgroup of the T$_d$ symmetry of the standard Mg$_{i}$ impurity, leads to a lifting of degeneracies \cite{Tinkham1964,Hamermesh1962} of excited states leading to the extra structure observed in these transitions. \par

\begin{figure}[h!]
\includegraphics[width=0.42\textwidth]{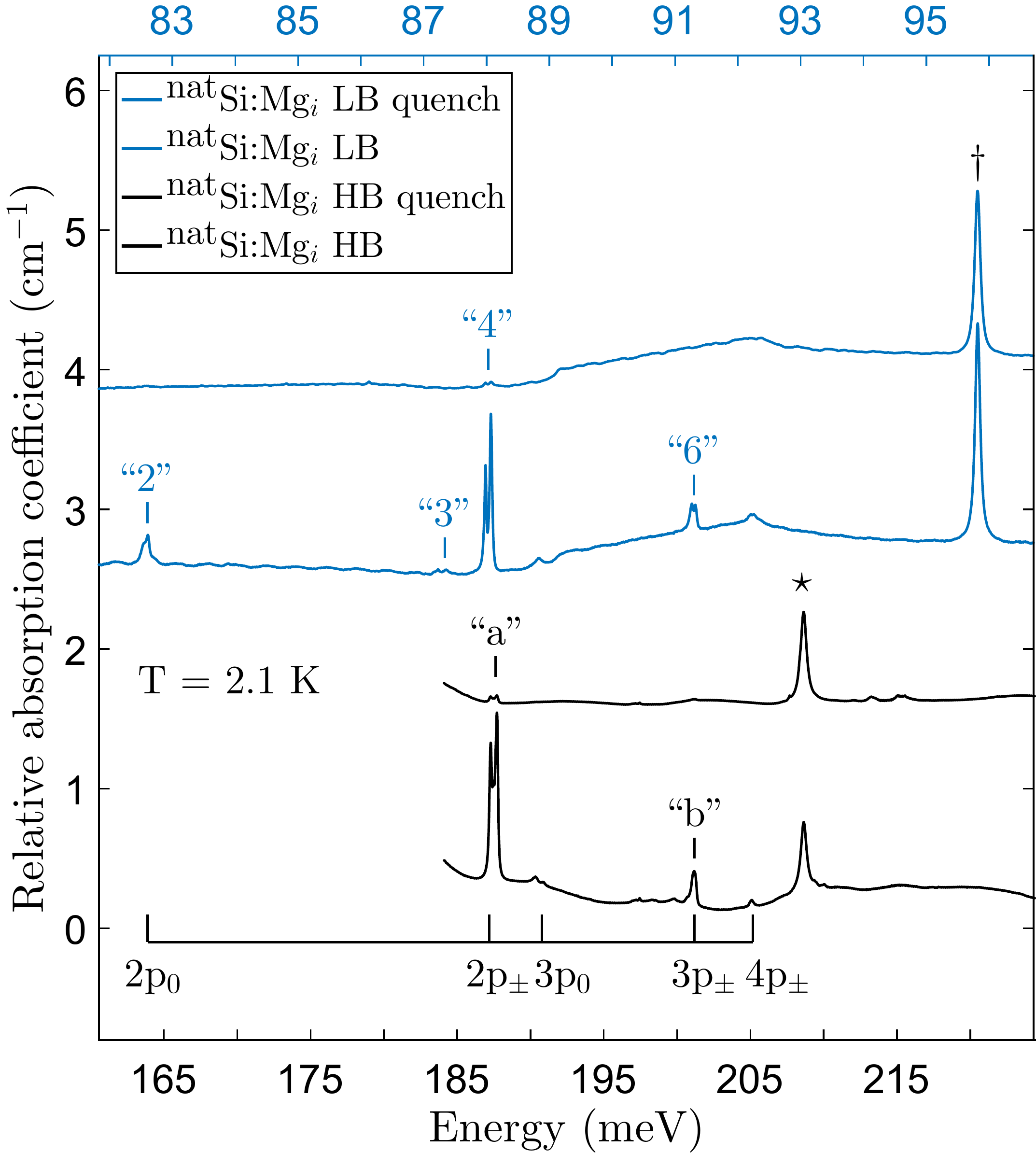}
\caption{Here we show the same samples viewed in absorption before and after a five minute $\SI{1200}{^{\circ}C}$ anneal/quench into alcohol. We have scaled the axes such that the ionized species are viewed over a range four times the width of the neutral donors such that the Mg$_{i*}$ transitions of both charge states are aligned on the plot. The top two spectra depict the neutral species and are plotted against the top axis. The bottom two spectra show the ionized species and correspond to the bottom axis. After the anneal/quench we see that Mg$_{i*}^0$ and Mg$_{i*}^+$ lines are reduced, if not eliminated altogether. For Mg$_{i*}^+$ the spectral region below 2p$_{\pm}$ is not included due to dewar window absorption. The $\dagger$ and $\star$ label the positions of the 2p$_0$ transitions of Mg$_{i}^0$ and Mg$_{i}^+$ respectively.}
\label{MgQuench}
\end{figure}

Another way to achieve a reduced symmetry configuration, namely the alternate interstitial symmetry site C$_{3v}$ along the $\langle$111$\rangle$ crystal axis, is discussed by Ho and Ramdas \cite{Ho1972} and Baxter and Ascarelli \cite{Baxter1973}. This predicts additional fine structure potentially consistent with what we observe in Mg$_{i*}^{0}$ and Mg$_{i*}^{+}$ spectra. However, as we note in Fig.~\ref{MgQuench}, a five minute anneal at $\SI{1200}{^{\circ}C}$ and subsequent quench into alcohol results in the reduction or elimination of all Mg$_{i*}$ lines. This points to the most likely scenario being that Mg$_{i*}$ is a complex that Mg forms with some other species. Such a center would dissociate at $\SI{1200}{^{\circ}C}$ and be prevented from reforming during the rapid cooldown of the quench. This behaviour would not be expected if Mg$_{i*}$ was due to an alternate interstitial site, since if anything, one would expect a more equal population of nearly degenerate interstitial sites at elevated temperature. We cannot make any definitive statements about the lifting of degeneracy we expect without knowing the precise nature of the complex. \par 

If the Mg$_{i*}$ center is a complex it seems likely that the second component is both ubiquitous and electronically neutral in silicon, since transitions of Mg$_{i*}$ have been seen in many samples \cite{2Ho2003,Ho2006} and Mg$_{i*}$ still acts as a deep double donor, just as it does when complexed with oxygen \cite{Ho2003}. With this in mind, carbon is a possible candidate. As noted by Jones et al. \cite{Jones1981}, substitutional acceptor complexes involving carbon referred to as X-centers are known to form among the Group IIIA elements. \par

\FloatBarrier
\clearpage

\subsection{Thermal effects}
We now discuss lines $``\textrm{1}"$, $``\textrm{3}"$, and $``\textrm{5}"$ previously observed by Ho and interpreted to be ground state to 2p$_0$, 2p$_{\pm}$, and 3p$_{\pm}$ transitions of a donor with an ionization energy $\SI{0.79}{meV}$ less than that of the donor responsible for lines $``\textrm{2}"$, $``\textrm{4}"$, and $``\textrm{6}"$ \cite{Ho2003,Ho2006}. We instead find that lines $``\textrm{1}"$ through $``\textrm{6}"$ all arise from the same perturbed Mg$_{i*}^0$ donor, with lines $``\textrm{2}"$, $``\textrm{4}"$, and $``\textrm{6}"$ originating from the ground state and lines $``\textrm{1}"$, $``\textrm{3}"$, and $``\textrm{5}"$ originating from very low-lying excited states which can be thermally populated even at liquid He temperatures. Peak positions for the resolved components of lines 1, 3, and 5 are included in Tab.~\ref{tab:ThermalActLineVals}. \par

\begin{figure}[htbp!]
\includegraphics[width=0.45\textwidth]{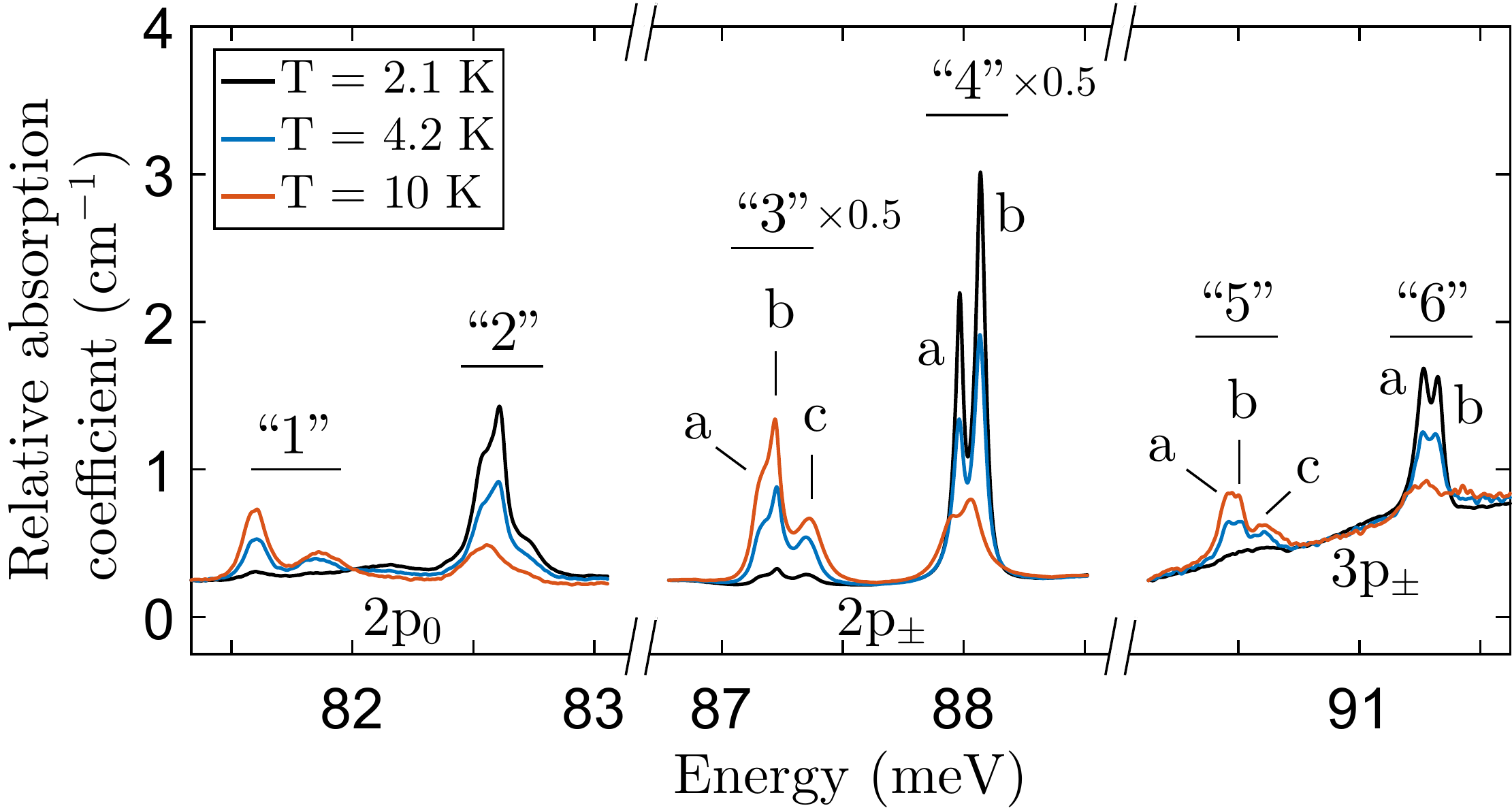}
\caption{Example spectra for three selected temperatures (2.1, 4.2, and $\SI{10}{K}$) showing thermal activation between Mg$_{i*}^0$ transitions referred to by Ho as lines $\textrm{``2"}$, $\textrm{``4"}$ and $\textrm{``6"}$ and the 2p$_0$, 2p$_{\pm}$ and 3p$_{\pm}$ of the thermally activated levels referred to as $``\textrm{1}"$, $``\textrm{3}"$, and $``\textrm{5}"$ \cite{Ho2003,Ho2006}. The p$_{\pm}$ levels of these are also labelled by visible components a-c, possibly indicative of two partially resolved doublets. Transitions are also labelled by Ho's assignments $\textrm{``1"}$-$\textrm{``6"}$. The 2p$_{\pm}$ region showing lines $\textrm{``3"}$ and $\textrm{``4"}$ has been scaled down by a factor of 0.5 for ease of viewing. Data shown corresponds to the $^{\textrm{nat}}$Si LB sample.}
\label{MgThermalLines}
\end{figure}

At pumped He temperatures the lines $``\textrm{1}"$, $``\textrm{3}"$, and $``\textrm{5}"$ intensities are very low, as already seen in Fig.\ref{Mg0star} where line $``\textrm{3}"$ can barely be observed at $\SI{2.1}{K}$, and lines $``\textrm{1}"$ and $``\textrm{5}"$ are unobservable. At $\SI{4.2}{K}$ the intensities of lines $``\textrm{1}"$, $``\textrm{3}"$, and $``\textrm{5}"$ are comparable to those of $``\textrm{2}"$, $``\textrm{4}"$, and $``\textrm{6}"$, and at $\SI{10}{K}$ the $``\textrm{1}"$, $``\textrm{3}"$, and $``\textrm{5}"$ lines are stronger than the $``\textrm{2}"$, $``\textrm{4}"$, and $``\textrm{6}"$ lines as seen in Fig.~\ref{MgThermalLines}. Focussing on the dominant 2p$_{\pm}$ transitions, we see that line $``\textrm{3}"$ has three components which we labelled 3a, 3b, and 3c. As already described, line $``\textrm{4}"$ is a doublet with components labelled 4a and 4b. As can be seen in Fig.~\ref{MgThermalLines}, lines 3a and 3b have very similar temperature dependence, while line 3c increases somewhat less with increasing temperature than do lines 3a and 3b. \par  

\begin{figure}[htbp!]
\includegraphics[width=0.42\textwidth]{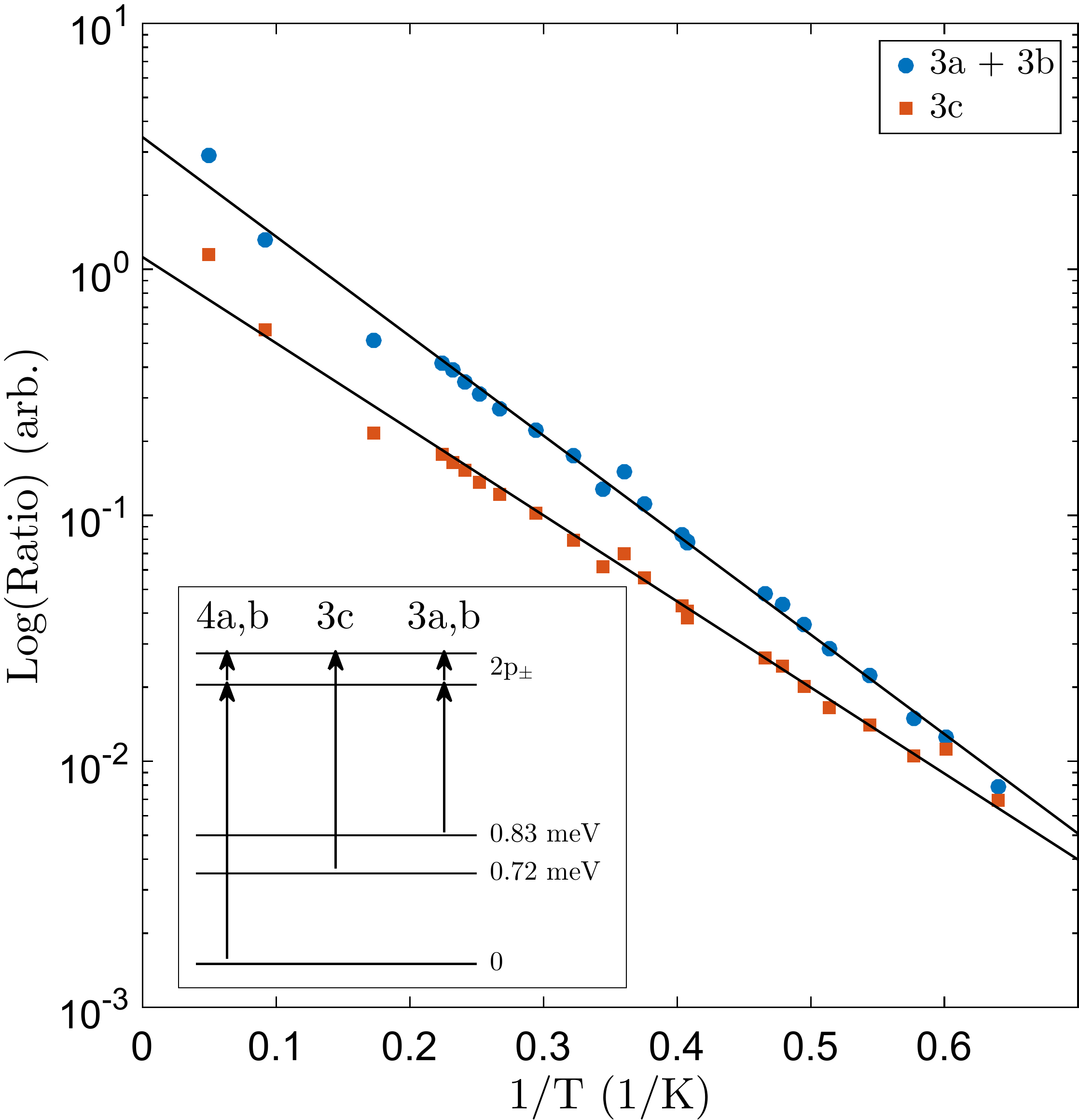}
\caption{Here we see an Arrhenius plot showing integrated intensities of Ho's line $``\textrm{3}"$ as a function of 1/T for components 3a + 3b together and line 3c. These are expressed as fractions of the integrated intensity of line $``\textrm{4}"$. In the inset we include a level diagram showing the main Mg$_{i*}^0$ 2p$_{\pm}$ transition along with transitions from two low lying levels $\SI{0.72 \pm 0.02}{meV}$ and $\SI{0.83 \pm 0.02}{meV}$ above the actual ground state. Here we consider the the positions of these levels to be the average of the optical spacings between components of of lines $``\textrm{3}"$ and $``\textrm{4}"$. These spacings match well to the thermal activation energies extracted from the slopes of these lines which are $\SI{0.80 \pm 0.02}{meV}$ and $\SI{0.69 \pm 0.03}{meV}$ for 3a + 3b and 3c respectively. Data shown corresponds to the $^{\textrm{nat}}$Si LB sample.}
\label{MgArrhenius}
\end{figure}

This is shown in more detail in Fig.~\ref{MgArrhenius}, an Arrhenius plot of the line 3a + 3b, and line 3c integrated intensity, vs. the line 4a + 4b integrated intensity as a function of inverse temperature. The integrated intensities of transition $\textrm{``3"}$ components were determined by fits to three mixed Gaussian-Lorentzian peaks, and those of transition $\textrm{``4"}$ to two peaks. The thermal activation energy of $\SI{0.80 \pm 0.02}{meV}$ for lines 3a + 3b, is in good agreement with the energy shifts between lines 3a (3b) and line 4a (4b), namely $\SI{0.82 \pm 0.02}{meV}$ ($\SI{0.84 \pm 0.02}{meV}$). The thermal activation energy of $\SI{0.69 \pm 0.03}{meV}$ for line 3c is similarly in good agreement with the energy difference of $\SI{0.72 \pm 0.02}{meV}$ between line 3c and line 4b. The origin of the line $``\textrm{3}"$ and line $``\textrm{4}"$ components is summarized in the inset to Fig.~\ref{MgArrhenius}. Very similar behaviour is observed for lines 5a, 5b, 5c, 6a, and 6b terminating in the 3p$_{\pm}$ levels. The behaviour of the line $``\textrm{1}"$ components vs. the $``\textrm{2}"$ components is also consistent with the low lying excited states shown in the inset of Fig.~\ref{MgArrhenius}. The greater splittings of the 2p$_0$ state, as already shown in Fig.~\ref{Mg0st_fineStruct} for line $``\textrm{2}"$, result in an overlap of transitions from the two low-lying excited states which cannot be resolved in line $``\textrm{1}"$. \par

The ground state structure of Mg$_{i*}^0$ is unusual, in that while the ground state is quite deep compared to the group V shallow donors, there are very low-lying excited states. For the group V donors the valley orbit excited states are more than $\SI{10}{meV}$ above the ground state. Attempts to observe similar low-lying excited states above the ground state of Mg$_{i*}^+$ center were unsuccessful. \par

\begin{table}[htbp!]
\begin{ruledtabular}
\begin{tabular}{l c}
Label &  \multicolumn{1}{c}{Transition energy (meV)}\\
\hline
$``\textrm{1}"$ (2p$_0$)             & \makecell[t]{81.56 \\ 81.60 \\ 81.87}\\
$``\textrm{3}"$ (2p$_{\pm}$) (a,b,c) & \makecell[t]{87.16 \\ 87.23 \\ 87.35}\\
$``\textrm{5}"$ (3p$_{\pm}$) (a,b,c) & \makecell[t]{90.46 \\ 90.51 \\ 90.61}\\
\end{tabular}
\end{ruledtabular}
\caption{Peak positions detailing fine structure of the thermally activated transitions labelled lines $``\textrm{1}"$, $``\textrm{3}"$, and $``\textrm{5}"$ in Fig.~\ref{MgThermalLines}. The values for line $``\textrm{3}"$ are given for spectra taken at $T$ = $\SI{2.1}{K}$ as with most other data presented in this work. For lines $``\textrm{1}"$ and $``\textrm{5}"$ the values correspond to data collected at $T$ = $\SI{10}{K}$ where the transitions were strong enough to observe and fit reliably.}
\label{tab:ThermalActLineVals}
\end{table}

\FloatBarrier
\subsection{New shallow donor (Mg-B)}
\begin{figure}[htbp!]
\includegraphics[width=0.42\textwidth]{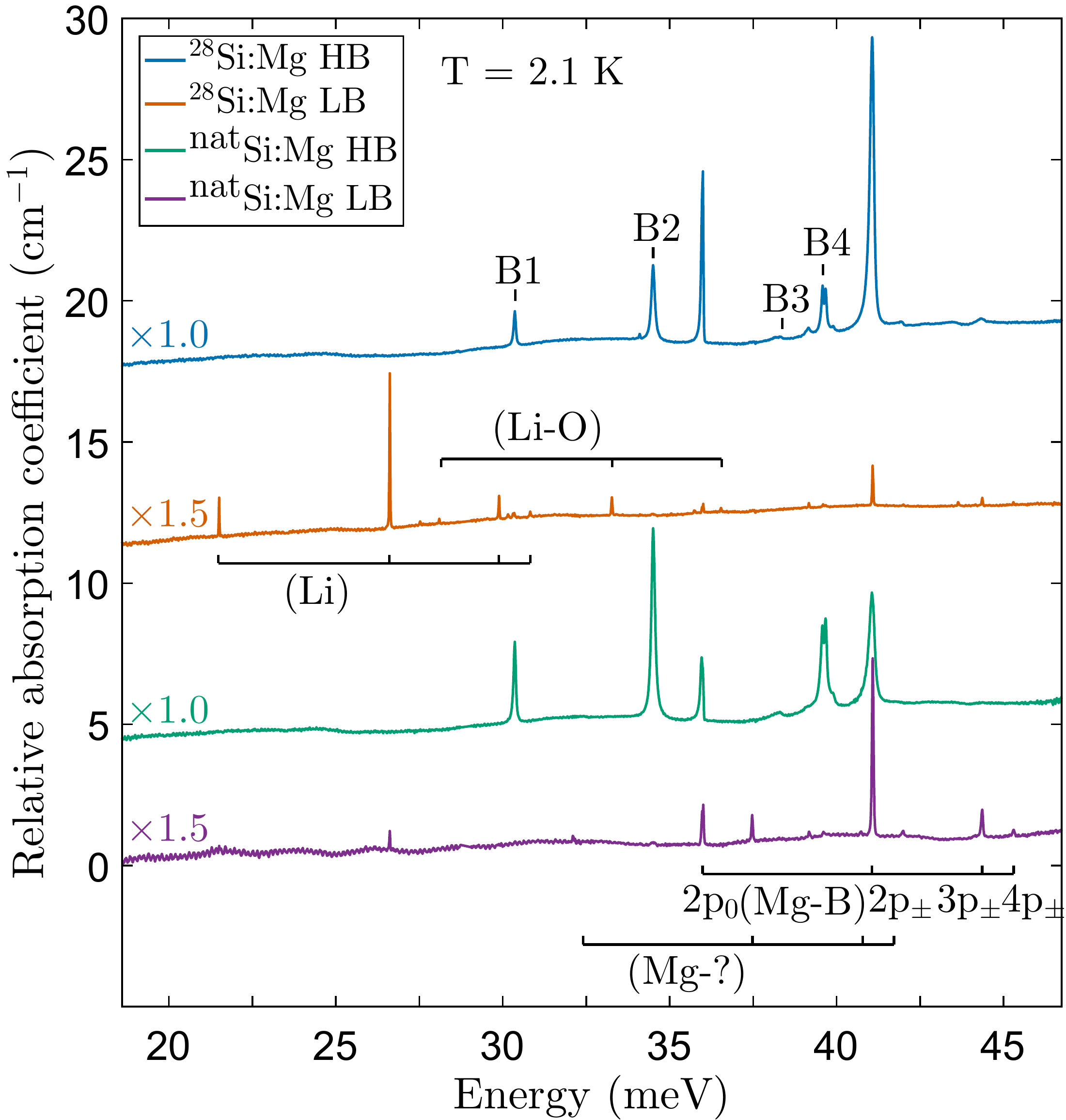}
\caption{Observation of the shallow donor region reveals a new center we identify as an Mg-B complex. In HB samples we observe strong boron acceptor features (B1 through B4 \cite{Pajot2010}) alongside the 2p$_0$, 2p$_{\pm}$, and, weakly, 3p$_{\pm}$ peaks of the Mg-B center. LB samples reveal less Stark broadened versions of these same lines up to 4p$_{\pm}$ along with other shallow donors including Li and Li-O and an unidentified Mg-? donor center.}
\label{shallow_donor}
\end{figure}

\begin{figure}[htbp!]
\includegraphics[width=0.42\textwidth]{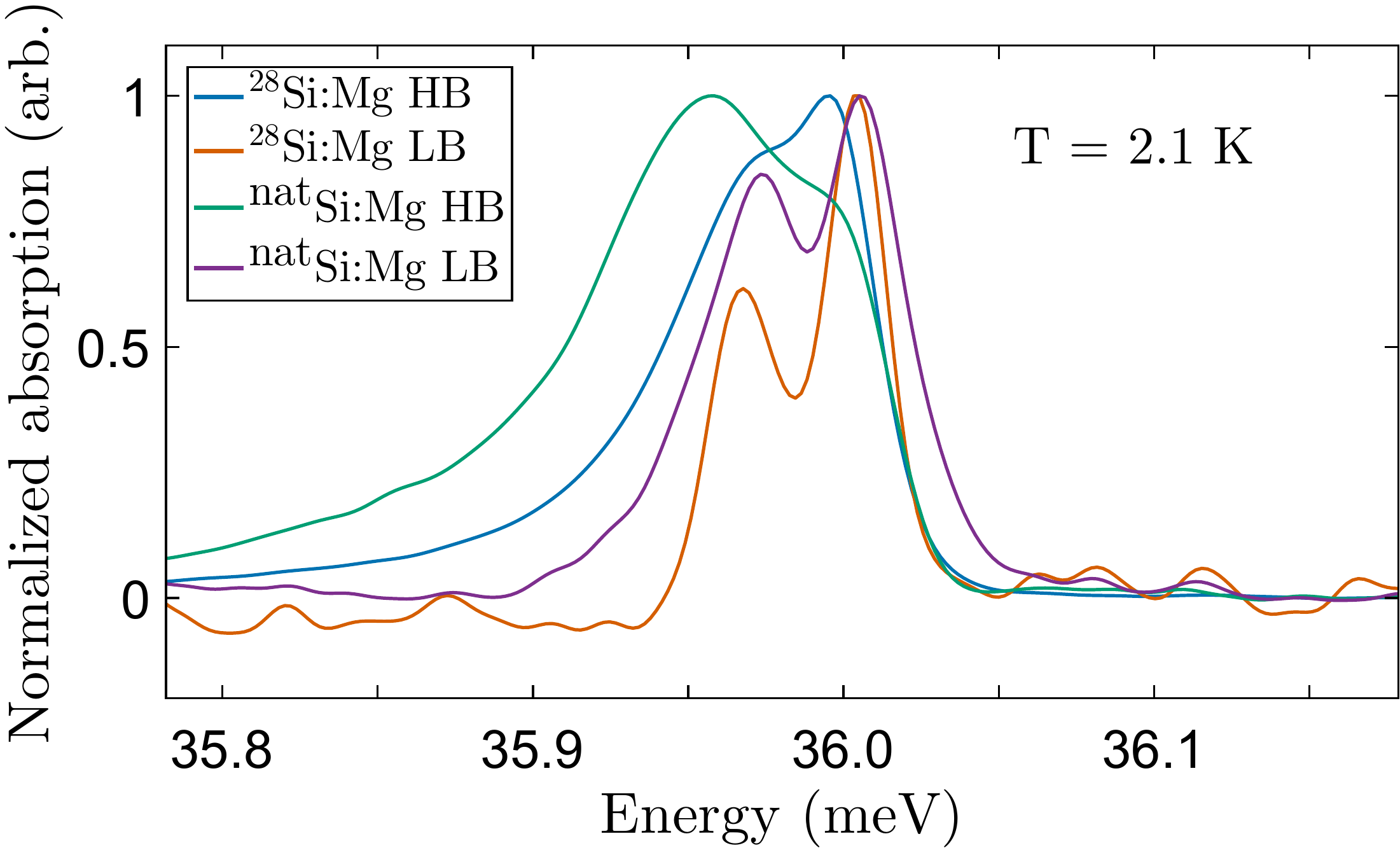}
\caption{On closer inspection we note in the shallow donor region that the 2p$_0$ transition of Mg-B is a doublet, while the 2p$_{\pm}$ transition and higher p$_{\pm}$ states are singlets.}
\label{MgBzoom}
\end{figure}

The presence of a new shallow donor we identify as a Mg-B pair is observed in the absorption spectra shown in Fig.~\ref{shallow_donor}. The presence of another shallow donor that may be Mg related, labelled Mg-?, is also noted. The HB samples reveal the Mg-B lines along with strong boron acceptor features. LB samples reveal higher excited states otherwise obscured by Stark broadening along with a number of other shallow donors. Fine structure in the 2p$_0$ line of the Mg-B complex which is revealed to be a doublet is visible in Fig.~\ref{MgBzoom}. Tab.~\ref{tab:shallowDonors} includes the peak positions of the excited states observed for Mg-B and Mg-? as seen in $^{\textrm{nat}}$Si. Some extra fine structure is not unexpected since, as a complex, Mg-B must by definition have reduced symmetry relative to the T$_d$ symmetry of the Si lattice. \par

The signature of Mg-B in Fig.~\ref{shallow_donor} is visible alongside those of a number of donor species, including Li and Li-O and a weakly visible Mg-? center. This last donor may represent magnesium associated with some acceptor other than boron \cite{Lin1982}. We estimate an ionization energy of $\SI{47.49}{meV}$ ($\SI{44.18}{meV}$) for Mg-B (Mg-?) via the energy of the highest visible transitions, namely 4p$_{\pm}$, and the corresponding theoretical binding energy for that state \cite{Pajot2010}. Energies of the Li and Li-O transitions visible primarily in the $^{\textrm{nat}}$Si spectrum of Fig.~\ref{shallow_donor} are tabulated by Pajot \cite{Pajot2010}. \par 

\begin{figure}[htbp!]
\includegraphics[width=0.42\textwidth]{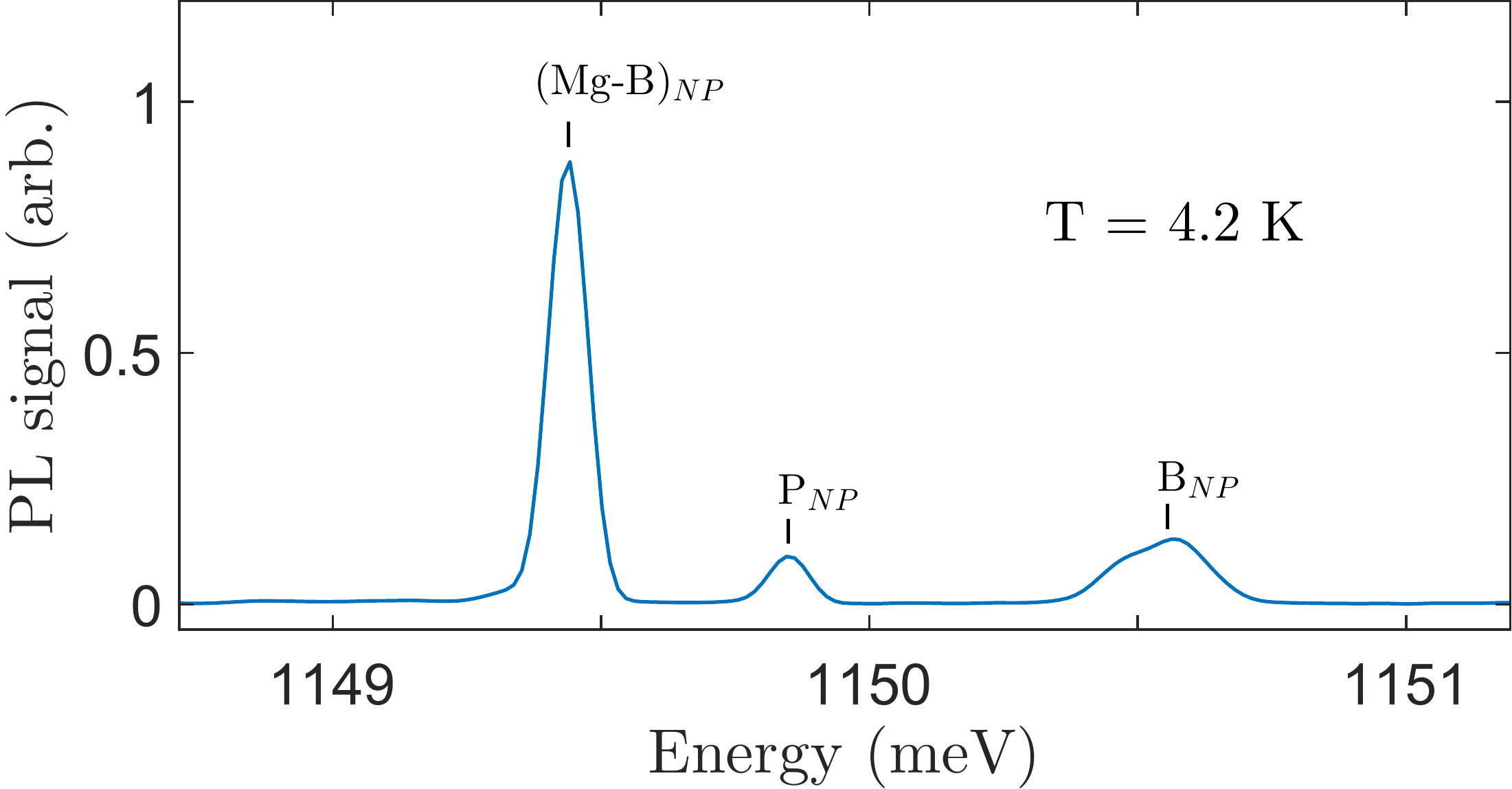}
\caption{Photoluminesence spectrum of the $^{28}$Si HB Mg-diffused sample under 200 mW of 532 nm illumination. The no-phonon (NP) peaks of phosphorus and boron are clearly visible together with a new peak we believe corresponds to the Mg-B donor bound exciton. Spectra were collected at $T$ = $\SI{4.2}{K}$ with $\SI{0.5}{cm^{-1}}$ ($\sim$ $\SI{0.062}{meV}$) resolution.}
\label{no_phonon}
\end{figure}

The Mg-B shallow donor is also observed in photoluminesence through the presence of a new donor bound exciton no-phonon line seen in Fig.~\ref{no_phonon}. We also observe the TA and TO phonon replicas of this feature visible together with bands of boron single and multi-exciton complexes (not shown). \par

\begin{table}[htbp!]
\begin{ruledtabular}
\begin{tabular}{l c c}
& \multicolumn{2}{c}{Transition energy (meV)} \\
\cline{2-3}
Label & Mg-B & Mg-? \\
\hline
2p$_0$     & \makecell[t]{35.97 \\ 36.01} & 32.10 \\
2p$_{\pm}$ & 41.08 & 37.47 \\
3p$_{\pm}$ & 44.36 & 40.73 \\
4p$_{\pm}$ & 45.30 & 41.99 \\
\end{tabular}
\end{ruledtabular}
\caption{Energies of donor transitions observed in the shallow donor region seen in Fig.~\ref{shallow_donor}. Values listed for Mg-B  and Mg-? are as seen in $^{\textrm{nat}}$Si.}
\label{tab:shallowDonors}
\end{table}

\begin{figure}[htbp!]
\includegraphics[width=0.42\textwidth]{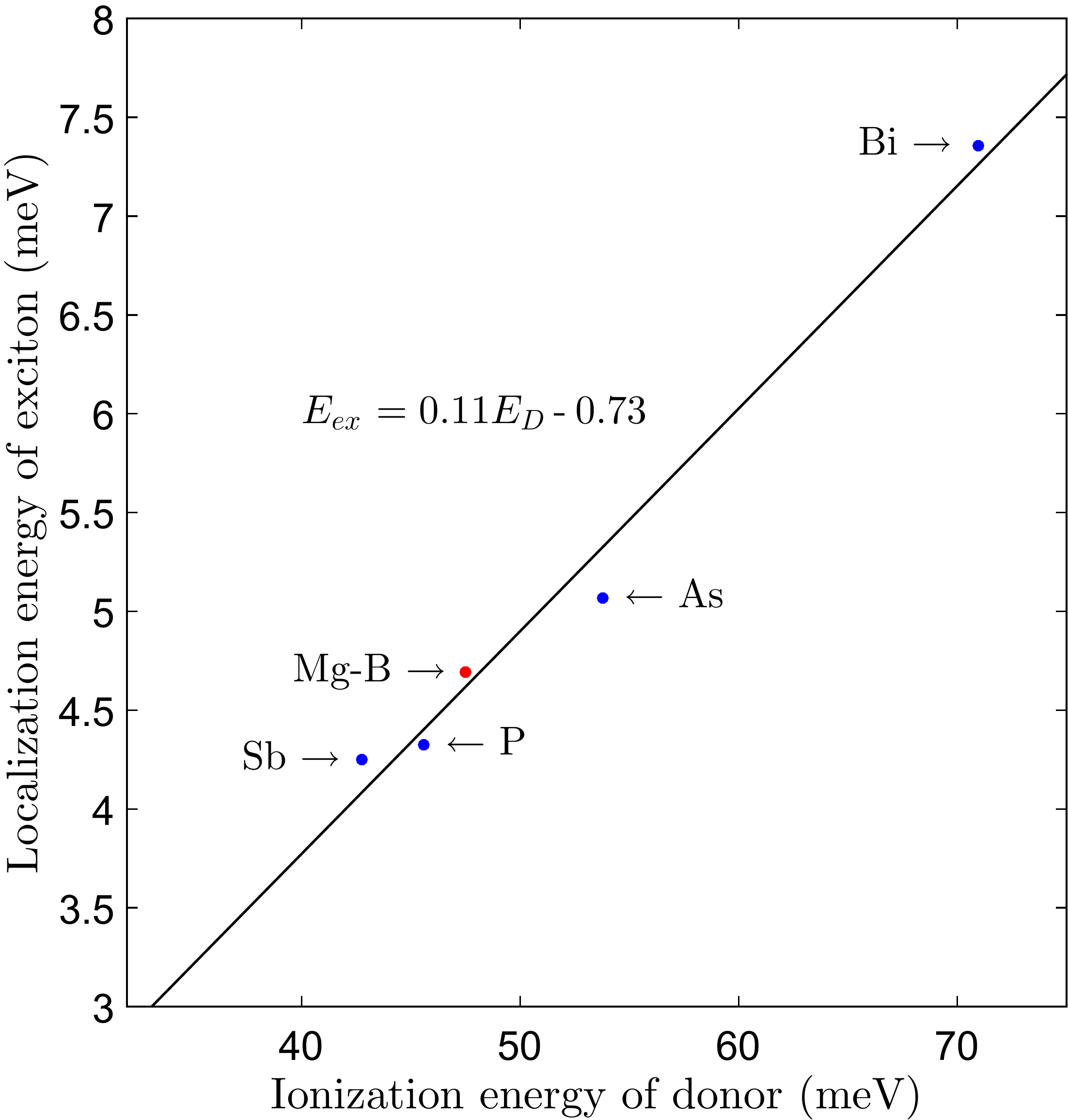}
\caption{Haynes' Rule demonstrating the linear relationship between the ionization energies and BE localization energies of several group-V donors and Mg-B.}
\label{Hayne}
\end{figure}

Haynes' rule \cite{Haynes1960} states that there exists a simple linear relation between the ionization energy of a donor/acceptor and the corresponding localization energy of the donor/acceptor bound exciton for that impurity. From the no-phonon region of our photoluminesence spectrum in Fig.~\ref{no_phonon} we are able to determine an estimated bound-exciton (BE) localization energy of $\SI{4.69}{meV}$ for Mg-B. With this and a number of other shallow donors studied we constructed the plot seen in Fig.~\ref{Hayne}, where we note that Mg-B follows the same trend. All our BE localization energies are determined relative to the phosphorus no-phonon BE feature visible in all samples. The phosphorus BE localization energy is calculated as the spacing between the phosphorus TA phonon replica and the low energy edge of the free-exciton TA yielding an estimate of 4.32 meV. \par

\FloatBarrier
\section{Conclusion and Outlook}
Our measurements have led to new interpretations for/discoveries of several features in magnesium-diffused silicon. A magnesium center previously observed in other work has been interpreted as Mg$_i$ in a reduced symmetry configuration, Mg$_{i*}$, either as a consequence of an alternate interstitial site or complexing with another impurity. Evidence currently lends weight to the latter possibility with an anneal/quench procedure showing signs of breaking up a likely Mg-related complex. Future studies will involve samples with relatively high carbon content to examine the impact on Mg$_{i*}$ centers. Stress studies to establish the symmetry of this center, allowing for a more quantitative discussion of the anticipated splitting of spectral features, could also follow. \par 

A new shallow donor center, observed for the first time in absorption, is identified as an Mg-B complex. Work to determine the origins of a different Mg-? shallow donor center by studying potential shallow donor complexes formed with other Group III acceptors such as gallium, indium, and aluminum is ongoing. The Mg-B complex is also seen in our photoluminesence results, revealing in emission the no-phonon/phonon replica transitions of the Mg-B donor BE. \par 

We were unable to see any of the 1s to 1s transitions in our current samples either for 1s(T$_2$) in absorption using FTIR spectroscopy or 1s(E) via Raman measurements (not shown). Further attempts to observe the 1s(T$_2$) and 1s(E) transitions of Mg$_{i}^+$ in absorption and Raman could be made in samples with optimized concentrations of Mg and B to allow observation of what are probably very weak features. \par

Further work using PL will follow, including a study of an isoelectronic feature noted earlier by Baber et al. \cite{Baber1988} and proposed to be due to a Mg-Mg pair center. \par

\FloatBarrier
\section{Acknowledgements}
This work was supported by the Natural Sciences and Engineering Research Council of Canada (NSERC). This work has been partly supported by the Russian Foundation for Basic Research (RFBF Project No. 18-502-12077-DFG) and of the Deutsche Forschungsgemeinschaft (DFG No. 389056032). The $^{28}$Si samples used in this study were prepared from Avo28 crystal produced by the International Avogadro Coordination (IAC) Project (2004-2011) in cooperation among the BIPM, the INRIM (Italy), the IRMM (EU), the NMIA (Australia), the NMIJ (Japan), the NPL (UK), and the PTB (Germany). \par

\end{document}